\documentclass[10pt,journal]{IEEEtran}

\renewcommand\arraystretch{1.8}

\usepackage{booktabs}
\usepackage{graphicx}
\usepackage{amsmath}
\usepackage{amssymb}
\usepackage{bbding}
\usepackage{enumerate}
\usepackage{stfloats}
\usepackage{tcolorbox}
\usepackage{multirow}
\usepackage{multicol}
\usepackage{subcaption}

\usepackage{tikz,xcolor,hyperref}

\definecolor{lime}{HTML}{A6CE39}
\DeclareRobustCommand{\orcidicon}{
	\begin{tikzpicture}
	\draw[lime, fill=lime] (0,0) 
	circle [radius=0.16] 
	node[white] {{\fontfamily{qag}\selectfont \tiny ID}};
	\draw[white, fill=white] (-0.0625,0.095) 
	circle [radius=0.007];
	\end{tikzpicture}
	\hspace{-2mm}
}

\ifCLASSOPTIONcompsoc
  \usepackage[nocompress]{cite}
\else
  \usepackage{cite}
\fi
\ifCLASSINFOpdf
\else
\fi
\begin{document}
%
\title{\textsc{StrTune}: Data Dependence-based Code Slicing for Binary Similarity Detection with Fine-tuned Representation}
%
%
%
%

\author{Kaiyan~He\!\href{https://orcid.org/0000-0002-9361-9093}{\orcidicon}, 
Yikun~Hu, Xuehui~Li, Yunhao~Song, Yubo~Zhao, Dawu~Gu
\IEEEcompsocitemizethanks{
\IEEEcompsocthanksitem Kaiyan He, Yikun Hu, Xuehui Li, Yunhao Song, Yubo Zhao, Dawu Gu are with the School of Electronic Information and Electrical Engineering, Shanghai Jiao Tong University, Shanghai, 200025, China (e-mail: pppgggyy@sjtu.edu.cn; yikunh@sjtu.edu.cn; tu.ana.ut@sjtu.edu.cn; songyunhao@sjtu.edu.cn; yubozhao@sjtu.edu.cn; dwgu@sjtu.edu.cn).

\protect\hfil\break
}
\thanks{(Corresponding Authors: Yikun Hu and Dawu Gu)}}

\IEEEtitleabstractindextext{%
\begin{abstract}
Binary Code Similarity Detection (BCSD) is significant for 
software security as it can address binary tasks such as malicious code snippets identification
and binary patch analysis by comparing code
patterns. Recently, there has been a growing focus on 
artificial intelligence-based approaches in BCSD due to their 
scalability and generalization.
Because binaries are compiled with different compilation configurations, existing approaches still face notable limitations when 
comparing binary similarity.
First, BCSD requires analysis on code behavior, 
and existing work claims to extract semantic, but actually 
still makes analysis in terms of syntax.
Second, directly extracting features from assembly sequences, 
existing work cannot address the issues of instruction reordering 
and different syntax expressions caused by various compilation configurations.

In this paper, we propose \textsc{StrTune}, which slices binary code
based on data dependence and perform slice-level fine-tuning.
To address the first limitation, \textsc{StrTune} performs backward slicing based on data dependence to capture how a value is computed along the execution. Each slice reflects the collecting semantics of the code, which is stable across different compilation configurations. \textsc{StrTune} introduces flow types to emphasize the independence of computations between slices, forming a graph representation. To overcome the second limitation, 
based on slices corresponding to the same value computation but having different syntax representation, \textsc{StrTune} utilizes
a Siamese Network to fine-tune such pairs, making their representations closer in the feature space. 
This allows the cross-graph attention to 
focus more on the matching of similar slices based on slice contents and flow types involved.
Our evaluation results 
demonstrate the effectiveness and practicality of \textsc{StrTune}. 
We show that \textsc{StrTune} outperforms the state-of-the-art methods 
for BCSD, achieving a Recall@1 that is 25.3\% and 22.2\% higher than 
jTrans and GMN in the task
of function retrieval cross optimization in x64.
\end{abstract}

\begin{IEEEkeywords}
Binary Code Similarity, Data dependence, Code Representation, Graph Neural Network
\end{IEEEkeywords}}

\maketitle

\IEEEdisplaynontitleabstractindextext

%
\IEEEpeerreviewmaketitle

%
%
%
%
%
%

\ifCLASSOPTIONcompsoc
\IEEEraisesectionheading{\section{Introduction}\label{sec:introduction}}
\else
\section{Introduction}
\label{sec:introduction}
\fi

%
%
%
%
\IEEEPARstart{B}{inary} Code Similarity Detection (BCSD) is commonly applied in retrieving
vulnerable functions in third-party libraries or firmware \cite{feng2016scalable, gao2018vulseeker, 
luo2023vulhawk, massarelli2019safe, xu2017neural, liu2018alphadiff, pewny2015cross, pewny2014leveraging, 
shirani2018b}, conduct malware analysis \cite{cesare2013control, farhadi2014binclone, hu2009large, jang2013towards} 
and perform binary patch analysis \cite{kargen2017towards, xu2017spain}. 
It is crucial for ensuring security as it identifies potential threats by analyzing and comparing binary code patterns. 
However, current work has not yielded satisfactory results in BCSD. Hence, it remains an issue to identify 
similar functions due to differences arising from cross-optimization, 
cross-architecture, and cross-compiler during binary compilation.


Recently, the field of artificial intelligence has witnessed 
significant advancements in BCSD. Existing AI-based research \cite{ding2019asm2vec, zuo2018neural, 
massarelli2019safe, guo2022exploring, ahn2022practical, wang2022jtrans, pei2020trex, feng2016scalable, xu2017neural, shalev2018binary, li2019graph, yu2020order, wang2023sem2vec, luo2023vulhawk} 
involves transforming the basic blocks into numerical 
vectors and using deep neural networks to generate function-level features. 
These methods can automatically extract features, 
have better generalization capabilities, and perform well at scale. However, 
they still suffer from low precision because directly embedding 
assembly instructions using NLP or graph neural networks 
fails to capture the features at the code behavior level. 
 
To identify similar functions across 
different compilation settings, 
current approaches typically have two limitations.
The first limitation is 
that BCSD needs to capture features of functions in terms of similar code behavior 
while existing approaches are unable to make an analysis on function behavior and 
actually extract features based on syntax and structures. 
These methods directly embed based on instruction tokens.
With a slight change at the instruction level that does not affect code behavior, 
the resulting embeddings may undergo significant alterations.
This is actually a feature at the syntax level and cannot 
reflect the code behavior of functions. \textsc{Pelican} \cite{zhang2023pelican} 
designs a trigger to insert some instruction into the binary 
code snippet, while preserving the semantics of programs. 
This also indicates that current methods tend to lean 
towards syntactic analysis rather than code behavior. 

The second limitation is that existing work 
has been insufficient in addressing the issues posed by instruction disorder and various 
syntax expressions resulting from different compilation configurations.
\textsc{Asm2Vec} \cite{ding2019asm2vec}, \textsc{Innereye} \cite{zuo2018neural} and 
\textsc{Safe} \cite{massarelli2019safe} all treat instructions within a single basic block 
or all sequential instructions as fixed sequences to extract features. 
\textsc{JTrans} \cite{wang2022jtrans} additionally 
considers the positional relationship of jump instructions.  
Different compilation configurations may 
lead to the disorder of instructions or even different syntactic forms
without affecting the computational content of function execution. 
However, these methods overly emphasize the positional association between instructions. 
Instructions without data dependence
can be interchanged forward or backward. These features 
of instructions make current methods of directly learning from instruction syntax not robust.
Also, instructions have numerous combination forms to accomplish 
the same value calculation, which cannot be covered by these methods.


For the aforementioned limitations, we propose 
a novel graph representation model and implement a prototype \textsc{StrTune} for BCSD. 
To address the first limitation, since the computed values represent the stable behavior of functions, instructions can be sliced based on data dependence, where each slice corresponds to the computation of a value in the function. \textsc{StrTune} 
performs backward instruction slicing based on data dependence and reorganizes them to 
generate slices as nodes.
Each slice represents the computation of one value, and such slicing ensures fixed instruction sequences within slices and data independence between slices. 
The computation of each value will invariably be manifested through different 
assembly instructions and sequences in similar functions and can be correspondingly captured by neural networks in the form of slices.

To tackle the second limitation, fine-tuning based on similar computational contents at the slice level can 
grasp identical slices under different syntactic expressions.
This helps the model pay more attention to similar slice pairs between functions. 
Since the instructions within a slice obtained through data dependence-based slicing 
have fixed positions, 
\textsc{StrTune} adds a new type of flow to represent the computational independence 
between slices, forming a graph representation for functions. Utilizing the 
RoBERTa natural language model, \textsc{StrTune} conducts a 
two-step operation of pre-training and fine-tuning. 
In the fine-tuning phase, \textsc{StrTune} pairs slices from the same line of computational contents in source program.
Then, \textsc{StrTune} uses a Siamese network to fine-tune slice pairs, allowing 
the cross-graph attention to focus more between similar slices based on slice contents and flow types involved. 

In this paper, we propose a novel pipeline \textsc{StrTune} for BCSD. 
First, we put forward a novel graph representation for binary
functions based on data dependence relationship. \textsc{StrTune} then employs a Siamese
network to fine-tune pairwise slices obtained from the same code. 
For graph-level representation learning, \textsc{StrTune} leverages the graph matching network to 
assign higher attention to semantically equivalent slices. Also, \textsc{StrTune} allows for 
visualization of the matching cross graphs, facilitating an understanding of the similarity calculation. 
To evaluate \textsc{StrTune}, we conduct function recalls in tasks cross-architecture, 
cross-optimization, and cross-compiler. \textsc{StrTune} outperforms state-of-the-art (SOTA) baselines in both recall@1 and MRR. 
Furthermore, we conduct a real-world vulnerability search. For each CVE, \textsc{StrTune} demonstrates an 
impressive recall success, ranking first in the most cases, surpassing other baselines by a considerable margin.

In summary, our contributions are summarized as follows:

\begin{itemize}
    \item We propose a novel graph representation model for binary code, slicing based on data dependence 
    which aids in segmenting instructions for value computation, and thus remains a 
    relatively resilient feature across various compilation configurations.
    
    \item We conduct a two-step of pre-training and pairwise fine-tuning for code slices 
    with the same computational content but different syntax, utilizing Siamese network 
    to make their representations closer.

    \item We employ the Graph Matching Network for function-level representation, with attention 
    coefficient focusing on the matching of similar slices. 

    \item We implement the above ideas and propose a model named \textsc{StrTune}. We validate 
    the effectiveness and practicality of \textsc{StrTune} by comparing it with SOTA methods. 
    Specifically, in the task of BCSD over different optimizations in x64, 
    \textsc{StrTune} achieves a Recall@1
    that is 25.3\%, 154.9\%, and 22.2\% higher than jTrans, VulHawk and GMN.

\end{itemize}

%
%
%
%
%
%

\section{BACKGROUND AND MOTIVATION}
\label{sec:background and motivation}

\subsection{Problem Description}

Binary code similarity detection (BCSD) is defined as the identification of similarities within 
assembly code when the source code is not available. 
We define two binary functions as semantically similar if 
they are compiled from the same or logically similar source code. 
This detection task is also viewed as a code search problem, 
where the aim is to locate the truly similar functions 
within a given pool of known functions for a query function. 
The challenge of binary code similarity highlights 
the necessity to abstract variations introduced by different 
compilers, compiler versions, optimization levels, architectures, 
and obfuscations. This is crucial for facilitating reverse 
engineering tasks, patch analysis, and the identification and 
remediation of binary vulnerabilities.

\subsection{Motivation}

This section explains the common problems in current work 
and our motivation for constructing a novel graph 
representation for binary functions.

\begin{figure}[!t]
  \centering
  \includegraphics[width=0.5\textwidth]{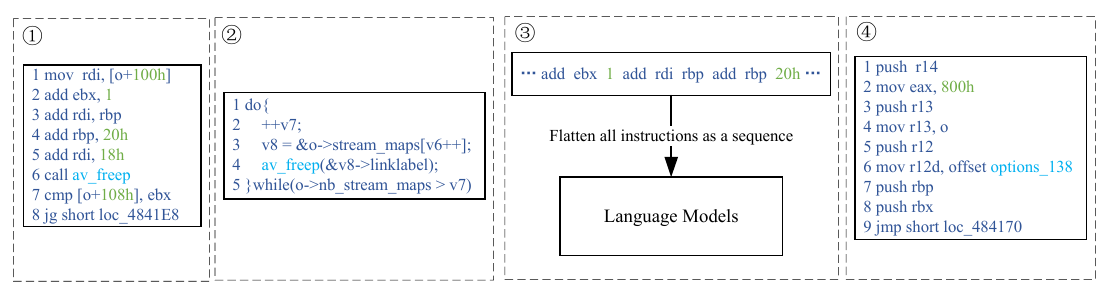}
  \caption{\textcircled{\raisebox{-0.9pt}{1}} shows instructions of the function 
  \textit{uninit\_options} with its corresponding computational contents in \textcircled{\raisebox{-0.9pt}{2}}.
  \textcircled{\raisebox{-0.9pt}{3}} shows the general method that current approaches use to process the instruction token.
  \textcircled{\raisebox{-0.9pt}{4}} involves the prologue of the function \textit{uninit\_options}.}
  \label{fig01}
\end{figure}

Figure \ref{fig01} captures a part of 
 the instructions from the \textit{uninit\_options} 
 function in the \textit{ffmpeg} binary, 
 compiled with the architecture x64, compiler gcc 
 and optimization O3. The instructions in \textcircled{\raisebox{-0.9pt}{1}} forms a complete basic block, 
 corresponding to the pseudocode in \textcircled{\raisebox{-0.9pt}{2}},
 with instructions in \textcircled{\raisebox{-0.9pt}{4}} showing the prologue of 
 this function. The current NLP-based approach, 
 as depicted in Figure \ref{fig01} \textcircled{\raisebox{-0.9pt}{3}}, involves 
 unfolding the instructions within a basic block 
 sequentially. 
 The entire sequence is considered as input to an NLP model for sentence-level learning. 
 This type of approach has several unreasonable aspects:

\begin{itemize}
\item In Figure \ref{fig01} \textcircled{\raisebox{-0.9pt}{1}}, 
\textit{`mov rdi, [o+100h]'} corresponds to the 
assignment of \textit{v8}, and \textit{`add ebx, 1'} corresponds 
to the increment of \textit{v7}. These two instructions 
can be interchanged without affecting the computation of two values. 
The instruction \textit{`add rbp, 20h'} represents the increment 
of \textit{v6} and must be placed after \textit{`mov rdi, [o+100h]'}. 
\textit{`add rdi, 18h'} is for parameter passing 
and must follow the \textit{`add rdi, rbp'} due to the 
data dependence of the register \textit{rdi}. Training 
based on NLP models would assume a fixed sequential 
relationship between these instructions. However, 
only instructions with data dependence 
have certain positional constraints, 
while instructions without data dependence can 
be arranged in any order without affecting the 
computational effect of the entire basic block. 
It is evident that the instructions in a basic block 
may contain several unrelated computation slices, 
whose sequences may be potentially changed by different 
compilation configurations.
\item In Figure \ref{fig01} \textcircled{\raisebox{-0.9pt}{4}}, instructions related 
to stack frame preservation for function calls 
are necessary for the function's runtime execution. 
However, some of these 
instructions, like \textit{`push rbp'}, may not be 
relevant to the computation of any values in this function. 
For instance, if \textit{rbp} is immediately 
assigned a variable in 
subsequent instructions, \textit{`push rbp' } 
becomes irrelevant to the overall 
semantic understanding of the function.
\end{itemize}

\begin{figure}[!t]
  \centering
  \includegraphics[width=0.48\textwidth]{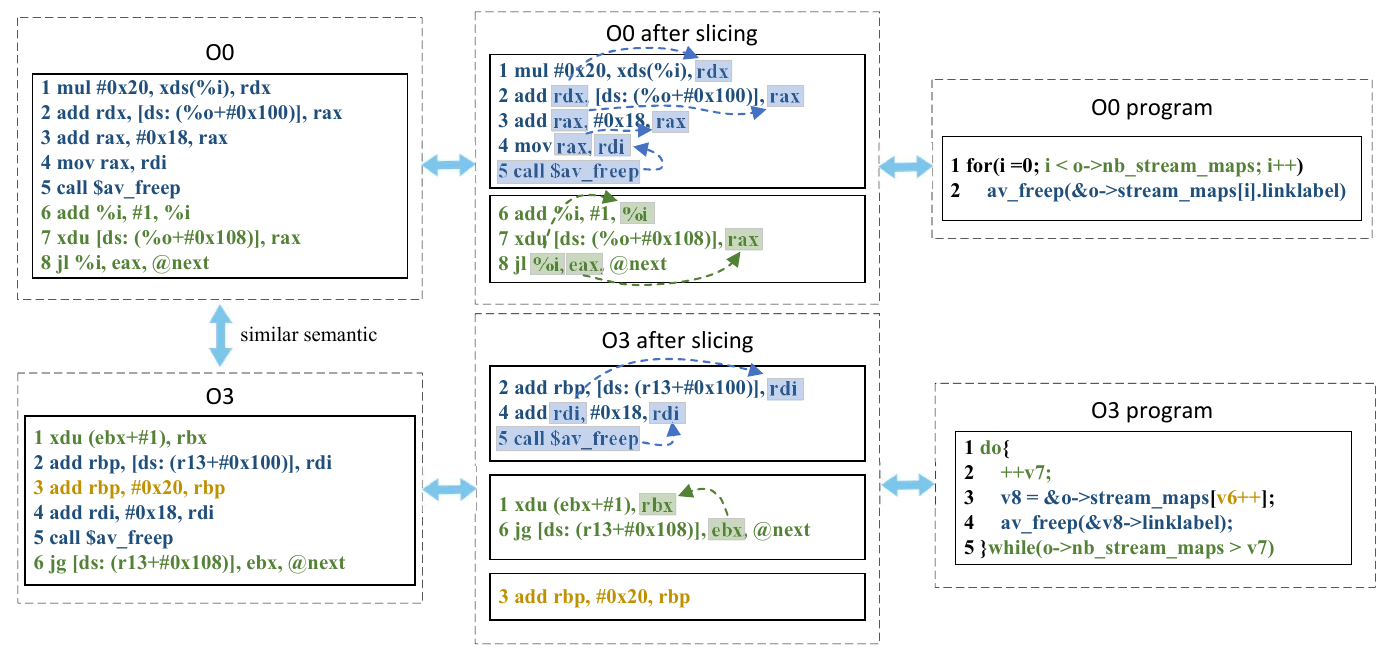}
  \caption{IR of the \textit{uninit\_options} function compiled with O0 and O3, and our ways to slice the instructions, corresponding to the computational contents in the same color.}
  \label{fig02}
\end{figure}

Computation refers to the value processing executed in a sequence of instructions, representing the concept of `collecting semantics’ \cite{cousot1992abstract} which captures and
records the changes in values throughout the execution of the code. Assembly code is compiled from source code, and although assembly instructions may change under different compilation conditions, the underlying computation of the source code remain unchanged, making it a stable feature. Therefore, this work mainly capture the computational semantics reflected in the assembly instructions for BCSD. Read and write operations of registers
in instructions are aimed at achieving these computations, 
which is reflected by data dependence. Hence, we perform backward slicing based on data dependence \cite{58784} to divide instructions into slices. Taking one basic block of O3 in Figure \ref{fig02} as an example, during backward traversal, Instruction 6 uses \textit{ebx}, and by tracing back to Instruction 1 where \textit{rbx} is defined, 1 and 6 form a slice. Next, for Instruction 5, it uses \textit{rdi} as the parameter of \textit{av\_freep}. Instruction 4 defines \textit{rdi} and uses \textit{rdi}, so the traversal continues until Instruction 2 defining \textit{rdi}. Thus, 2, 4, and 5 form another slice. Finally, Instruction 3, which has no preceding data dependence, forms its own slice.

Slices obtained from the same basic block are data independent, 
and the contents within each slice correspond to 
computations of certain values in the program. 
We consider slices performing 
equivalent computations on the same values in the 
program as semantically similar. 
Therefore, slices that are 
semantically similar but syntactically different 
under various compilation configurations can be 
matched one-to-one after being split in this way.

As an example, we consider the similar IR 
slices of the \textit{uninit\_options} function compiled 
with O0 and O3 respectively, as shown in Figure \ref{fig02}. 
The slices based on data dependence correspond
to the function calls and conditional statements respectively. 
We pair slices with the same color after pre-processing to fine-tune the model, 
aiming to learn instructions that have the same semantic 
but different syntactic forms. 
During slices matching, instructions in yellow will not be matched with any slices compiled with O0.
Slices in yellow represent the increment operation of variable \textit{v6} introduced in O3 optimization.
The role of variable \textit{v6} is equivalent to that 
of variable \textit{v7}. This is because different compilation configurations may trigger optimizations, leading to the generation of additional intermediate computation variables in the assembly code. However, we do not match slice pairs for these intermediate variables based on slice matching. Therefore, during later-stage feature extraction, the neural network assigns lower attention to these slices, aiding in the accurate similarity calculation.

It's important to consider the connection relationships between slices, as the structure training takes into account the contents of slices and the types of edges involved. Therefore, We add four types of edges to the graph based on whether there is data dependence (\textcircled{\raisebox{-0.9pt}{2}} Data Parallel Flow, \textcircled{\raisebox{-0.9pt}{3}} Data Dependence Flow) or control flow dependence (\textcircled{\raisebox{-0.9pt}{1}} Sequential Flow, \textcircled{\raisebox{-0.9pt}{4}} Jump Flow), the model can focus on the flow process of variables in the graph, thereby highlighting the more similar parts between graphs.

\begin{figure}[!t]
  \centering
  \includegraphics[width=0.48\textwidth]{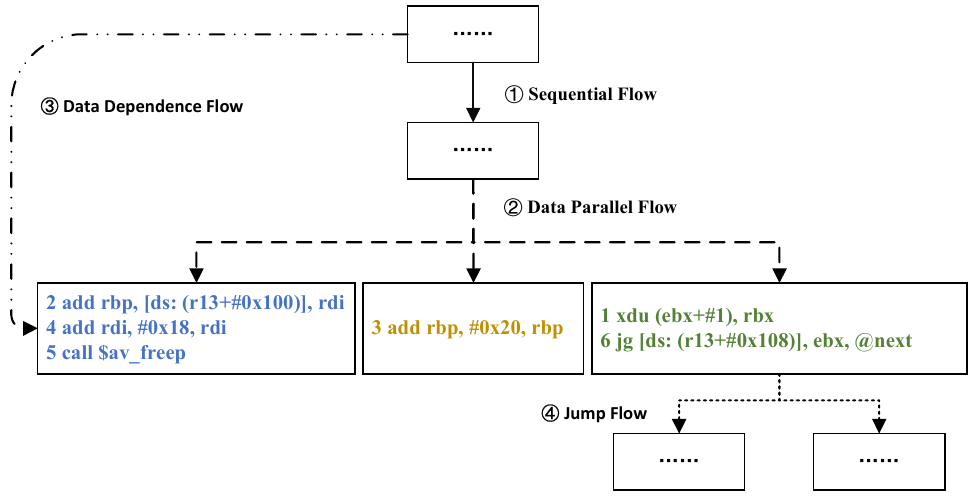}
  \caption{Example of our graph representation, consisting of four types of flow: 
  \textcircled{\raisebox{-0.9pt}{1}} Sequential flow, \textcircled{\raisebox{-0.9pt}{2}} Data parallel flow, 
  \textcircled{\raisebox{-0.9pt}{3}} Data dependence flow and \textcircled{\raisebox{-0.9pt}{4}} Jump flow.}
  \label{fig03}
\end{figure}

As shown in Figure \ref{fig03}, after the previous step, we get our slices to serve as our graph nodes. 
Supposing the slices $s_{1}$ and $s_{2}$ respectively belong to the basic block $bb_{1}$ and $bb_{2}$, 
we add flow type between slices according to the following rules:

\begin{enumerate}
\item Sequential flow: To preserve the original unconditional jump information. If $s_{1}$ is equal to $bb_{1}$, and $s_{2}$ is equal to $bb_{2}$, 
meaning both basic blocks are split into only one slice each. Also, there is 
an unconditional jump between $bb_{1}$ and $bb_{2}$, we add a sequential flow between $s_{1}$ and $s_{2}$. 
\item Jump flow: To represent the original conditional jump. If both basic blocks are split into only one slice each, and there is a conditional jump 
between $bb_{1}$ and $bb_{2}$, we add a jump flow between $s_{1}$ and $s_{2}$. 
\item Data dependence flow: To capture the long-distance data dependence.
If there exists data dependence between $s_{1}$ and $s_{2}$, 
meaning a variable defined in $s_{1}$ is used in $s_{2}$, we add 
a Data dependence flow between $s_{1}$ and $s_{2}$. It is worth 
noting that this type of flow only occurs between slices split 
from different basic blocks because our slicing already implies 
the data dependence within basic blocks.
\item Data parallel flow: Focusing on the unfixed
positions between slices without data dependence. This type of flow can be further divided into jump parallel and sequential parallel. 
Taking sequential parallel as an example. If there is 
an unconditional jump between $bb_{1}$ and $bb_{2}$ and $bb_{1}$ or $bb_{2}$ is split into more than one slice,
we add a sequential parallel flow between $s_{1}$ and $s_{2}$. The same is for the addition of jump parallel flow.
  Data parallel flow emphasizes 
the unfixed positions between slices, facilitating slice alignment between functions.
\end{enumerate}

Therefore, slicing based on data dependence satisfies three features: 1. Instructions within a slice have data dependence and thus have fixed positions and cannot be swapped. 2. Slices split from a basic block do not contain data dependence, and their computations do not affect each other, hence we consider them as slices with data parallel. 3. A slice corresponds to the computation of a certain value in the source code shown in the middle column of Figure \ref{fig02} with the same color. According to this pattern, we can obtain pairs of slices with different syntax but the same computational content, enabling the natural language model to generate similar embeddings for slices with similar semantic but different syntax. Additionally, in structural perception, the network can identify similar slices (comprising node contents and related edges). The greater the presence of such similar features, the higher the similarity of the corresponding source code computations. Consequently, the model tends to regard such pairs of functions as more similar.

\section{DESIGN}
\label{sec:design}

\begin{figure*}[!t]
  \centering
  \includegraphics[width=0.98\textwidth]{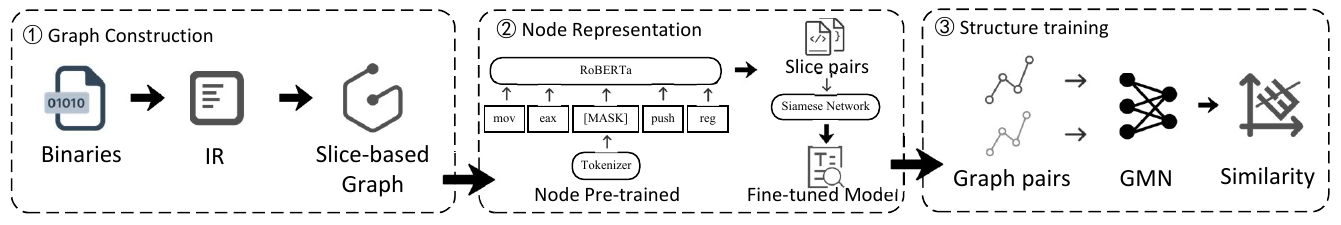}
  \caption{The overview of \textsc{StrTune} consisting of three components: \textcircled{\raisebox{-0.9pt}{1}} Graph Construction 
  that builds novel graphs base on IR lifted from binaries. \textcircled{\raisebox{-0.9pt}{2}} 
  Node Representation that forms a two-step learning process for similar slices, including pre-training and fine-tuning 
  of RoBERTa model. \textcircled{\raisebox{-0.9pt}{3}} Structure training that performs labeled learning for graph pairs based on the 
  node representations given by RoBERTa, utilizing attention coefficients cross graphs. \textsc{StrTune} forms
  a binary similarity model which takes two functions as input and outputs a similarity score.}
  \label{fig04}
\end{figure*}

\subsection{Overview}

Figure \ref{fig04} illustrates the entire workflow of our approach, 
which primarily consists of three components. 

\textbf{Graph Construction (\textcircled{\raisebox{-0.9pt}{1}}).} We lift the binary code into Intermediate Representation (IR)
Microcode and perform instruction deletion and preservation due to specific rules as the pre-processing.
Based on data dependence, we apply backward slicing to decompose the instructions into slices at 
the basic block level, as nodes of our graph. Each node 
can independently correspond to the computation of some values.
Also, we categorize the edges into four types and thus form a representation graph for functions. 

\textbf{Node Representation (\textcircled{\raisebox{-0.9pt}{2}}).} Based on the obtained graph, we normalize the instructions 
within nodes and perform a two-step operation involving 
pre-training and fine-tuning using a natural 
language model RoBERTa to obtain slice representations. We form pairs of code slices corresponding 
to the same computation content and use a Siamese Network for contrastive learning. 
This approach brings the embeddings of slices with different syntax but same semantic closer, 
facilitating the learning of code slices across different compilation configurations.

\textbf{Structure training (\textcircled{\raisebox{-0.9pt}{3}}).} Based on slice embeddings and different 
flow types, we employ Graph Matching Network (GMN) 
between a pair of graphs, which can assign higher attention coefficients to similar slices across graphs.
The network obtains two graph-level embeddings and calculates
similarity scores. Finally, \textsc{StrTune} takes pairs of 
functions as inputs and outputs
similarity scores.

\subsection{Graph Construction}

\begin{figure}[!t]
  \centering
  \includegraphics[width=0.48\textwidth]{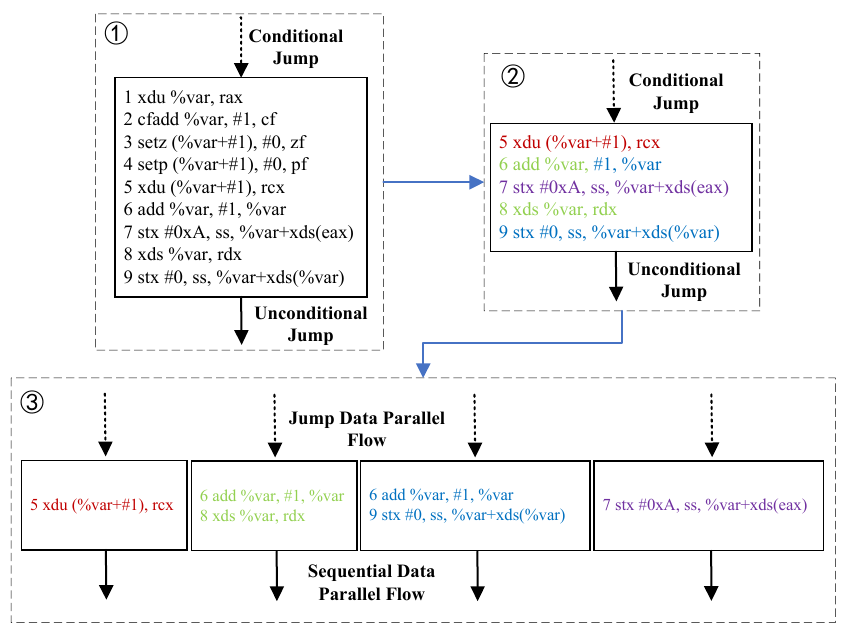}
  \caption{Graph Construction of our model \textsc{StrTune}: \textcircled{\raisebox{-0.9pt}{1}} shows unprocessed Microcode in a 
  basic block. After removal and preservation, we obtain the remaining instructions in \textcircled{\raisebox{-0.9pt}{2}}, 
  where instructions of the same color exhibiting data dependence, are considered as one code slice. Then we add specific 
  flow type between slices based on the rules mentioned before, as shown in \textcircled{\raisebox{-0.9pt}{3}}.}
  \label{fig05}
\end{figure}

We first lift the binary to Microcode 
for analysis. Figure \ref{fig05} \textcircled{\raisebox{-0.9pt}{1}} shows a segment of Microcode corresponding 
to a real-world program. Each Microcode instruction consists of one opcode and 3 operands: 
left, right, and destination, although some 
operands may be missing for certain types of instructions. 
We first perform pre-processing of Microcode based on following rules.

\textbf{Removal.} Since Microcode includes EFLAGs 
as implicit operands, considering all EFLAGs 
introduces extra overhead and obscures the 
main feature of the binary functions. Therefore, 
we retain only those EFLAGs assignment instructions 
where the defined EFLAGs are used in subsequent blocks. 
As shown in Figure \ref{fig05} \textcircled{\raisebox{-0.9pt}{1}}, 
\textit{cfadd} generates the carry bit. \textit{setz} and \textit{setp} compare the 
left and right operands and store the results 
in the \textit{zf} and \textit{pf} respectively. 
These EFLAGs are either redefined or 
not used in subsequent instructions. Additionally, 
\textit{rax} defined in the first instruction 
is also redefined before used, so the 
assignments to EFLAGs and registers in these situations 
can be removed without influencing the computation of values 
in the functions.

\textbf{Preservation.} We also preserve unused registers 
in two cases. The first case is for assignments to 
global variables, identified as mop\_v in Microcode. 
The second case is for passing arguments to a called 
function. Since the registers used for passing arguments 
differ for each architecture, and because the default 
function call is immediately followed by the argument 
passing instruction within the same basic block, we 
preserve instructions that assign values to unused 
registers before a call instruction. If the register 
is an argument register for the corresponding architecture, 
we also retain that instruction. Therefore, we pre-process the instructions in 
Figure \ref{fig05} \textcircled{\raisebox{-0.9pt}{1}}
and obtain Figure \ref{fig05} \textcircled{\raisebox{-0.9pt}{2}}.

\textbf{Slicing and Flow Adding.} Based on the data dependence, we perform backward slicing and grouping of the instructions as mentioned in 
Section \ref{sec:background and motivation}. Starting from Instruction 9, it uses \textit{var}, which is defined by Instruction 6, so 6 and 9 form a slice. Instruction 8 also uses \textit{var}, and since one instruction can be associated with multiple slices, Instructions 6 and 8 form another slice. The remaining Instructions 7 and 5 each form their own slice. These obtained slices are considered data-parallel segments, having unfixed position with each other but fixed sequence of instructions within themselves. Therefore, even if a slice is produced by different compilation options, as long as it corresponds to the same computational logic in the source code, it can be captured and aligned by the subsequent network. Additionally, since the basic block is connected to the previous one by a conditional jump edge, all the slices are linked with jump data parallel flow coming from the previous slices. When encountering a conditional jump as an outgoing edge in the basic block, we only connect the slice containing the jump instruction with the subsequent slices. When encountering an unconditional jump, as in this example, we add sequential data parallel flow from each slice to the subsequent slices, forming the graph as shown in Figure \ref{fig05} \textcircled{\raisebox{-0.9pt}{3}}. At the same time, based on the data definition and use across basic blocks, these slices are also connected from previous slices with data dependence flow to handle remote data dependence. On the basis of slice alignment, by matching the flow types connecting the slices and the content of the connected slices, the network can enhance the confidence of similar slices using attention coefficient, thereby increasing the similarity score of similar functions.

Based on our observations, different optimization levels can lead to duplication of the same slice, 
resulting in repeated slice with the same semantic in the graph. 
We merge nodes that are completely identical in content 
and have the same preceding and following nodes, as well as the same types of connecting edges, 
achieving a partial effect of manual de-optimization. 
Additionally, we add data dependence flows between code slices.
Since the splitting of slices already reflects the intra-block data dependence chain, 
data dependence flows only occur between slices split from different basic blocks.

\subsection{Node Representation}

\begin{figure}[!t]
  \centering
  \includegraphics[width=0.48\textwidth]{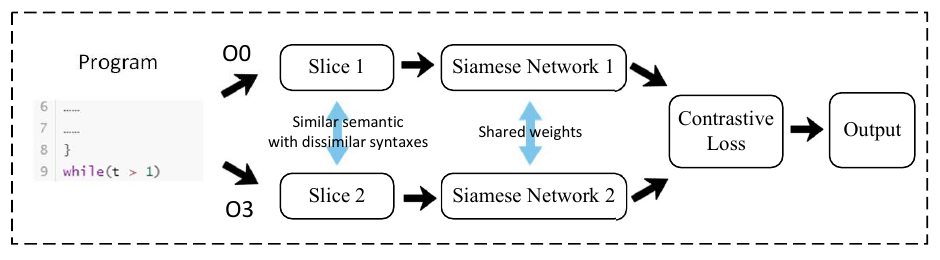}
  \caption{Siamese neural network for fine-tuning RoBERTa model on similar slices with identical computation contents but different syntax.}
  \label{fig06}
\end{figure}

In this section, we demonstrate how to learn the content of nodes in the constructed graph. 
Figure \ref{fig06} shows the pipeline we use for slice content learning.

\textbf{Tokenization and Normalization:} As previously mentioned, 
a Microcode instruction is divided into one opcode and three operands, 
providing a natural condition for our tokenization. Since the opcode 
determines the basic type of an instruction, such as movement, shift, 
comparison, etc., it provides fundamental semantic features.
Given the limited number of opcodes in Microcode, we construct 
a separate vocabulary for opcodes to alleviate out-of-vocabulary (OOV) 
issues during training and avoid fundamental semantic loss. Due 
to the chosen maturity level of Microcode parsing, operands 
might appear in compound expressions after some transformations, 
such as \textit{var\_160+\#1}, which increases the complexity 
of operands. Forms with an underscore suffix typically involve 
the definition and use of a temporary variable. Since our 
instruction slices already consider such local information, 
and this global information is also connected through data 
dependence flows between slices, we directly remove the suffix 
such as `\_160'. 
Also, terms starting with `\#' are usually 
numeric constants, which we replace with the \textit{num} 
character. String constants, usually carrying richer 
semantic, are retained in their entirety. 
Fortunately, microcode categorizes operands, such as mop\_S for 
local stack variables and mop\_l for local variables. 
Therefore, we can replace OOV operands with their types, 
minimizing the semantic loss in instruction representation learning.

\textbf{Pre-training:} We use RoBERTa, a natural language 
processing model based on the Transformer architecture, 
to perform pre-training on our normalized code slices. 
RoBERTa is designed for unsupervised training. Given a slice, 
we randomly select 15\% of 
the opcode or operand tokens and replace the chosen words 
with a special [MASK] token. The model's objective is to 
predict the word marked by [MASK] based on the context.

We collect all slices from the 
training dataset to form our pre-training model 
and optimize the model parameters using cross-entropy:

\begin{equation}
L\left( \theta  \right) =  - \sum\limits_{i \in M} {\log P\left( {{x_m}|{x_{ < m}};\theta } \right)} 
\end{equation}
where $M$ represents the mask set, $x_m$ represents the original token 
marked as [MASK], $x_{<m}$ represents the context tokens of that 
mask token, and $\theta$ represents the model parameters. 
The model, after pre-training, can transform a given slice into an embedding vector in a 
high-dimensional vector space, where each dimension corresponds 
to different aspects of semantic information. We believe 
this embedding initially extracts the information from 
the slices but lacks targeted learning 
for instruction slices with the same semantic but different syntax.

\textbf{Fine-tuning:} To enhance the RoBERTa model's understanding 
of slices with the same computation content, we constructed a Siamese network. 
This network improves the quality of embeddings by matching 
slices from the same source code
but compiled under different configurations. For example, given the 
significant optimization differences between O0 and O3, 
we primarily select slices from these two optimization 
levels for pairing, keeping the compiler, bit, and 
architecture unchanged. We can obtain two binaries compiled 
at O0 and O3 optimization levels as the objects to be matched.

Specifically, for a given C source code file, we consider 
that code involving conditional judgments (if, while, for loops) 
often contains more semantic information, which tends to be stable 
across different compilation configurations. Branch statements execute different code blocks based on input conditions, determining the behavior of the program. Instructions involving branch statements from different compilations are guaranteed to exist and be semantically equivalent. Therefore, we select statements involving conditional judgments and randomly pick 10\% of statements without conditional jump to form our input set from this source file as the appropriate size of training dataset. For each program statement, 
we perform address mapping 
in the binary to obtain pairs of slices $(s_{1}, s_{2})$ under O0 and O3, representing slices with the same computation content in respect of different optimizations.
We also construct slice pairs for different architectures, compilers, and bitness to 
enable the model to learn slice representations more comprehensively.

We utilize contrastive loss as the loss function:

\begin{equation}
L = \sum {y{d^2}}  + (1 - y)\max {(m - d,0)^2}
\end{equation}
where $y = 1$ indicates that the selected pair of slices are semantically similar, while $y = 0$ indicates dissimilarity.  
$m$ represents a threshold, where if the distance exceeds this threshold, dissimilar sample pairs will no longer incur additional loss. This hyper-parameter is set to 1 in our experiments. $d$ denotes the cosine distance of two vectors $\frac{{{v_1} \cdot {v_2}}}{{\left| {{v_1}} \right| \times \left| {{v_2}} \right|}}$ generated by our pre-trained Roberta model. The model utilizes the cosine similarity of vectors of slice pairs to ensure that the pre-trained Roberta model maps slices corresponding to the same statement as close as possible, while pushing semantically dissimilar slices farther apart. Through this fine-tuning step, the Roberta model is capable of generating robust embeddings when faced with semantically similar slices under different compilation conditions.

\subsection{Structure Training}

We use Graph Matching Networks (GMN) as the structure-level feature extraction model. With respect to graph neural networks, GMNs not only consider aggregated messages on individual graphs but also 
incorporate a cross-graph attention mechanism. This attention 
coefficient measures the degree of alignment between cross-graph nodes. 
Cross-graph attention matches well with 
our fine-tuning based on code slices.

Given a pair of functions to compare, the node features 
are embeddings of slices generated 
by the RoBERTa model. Considering that our constructed 
function representation graph includes five types of edges, 
we adopt the commonly used encoding technique in machine 
learning and data processing, one-hot encoding, to distinguish 
the types of flow in the graph. The GMN model consists of 
an encoder, propagation layers, and an aggregator. The encoder 
maps the node and edge features to initial node and edge vectors 
through separate MLPs (Multi-Layer Perceptrons). Then, through 
multiple layers of propagation, the representation for each 
node will accumulate information in its local neighborhood 
and edges. Specifically, the propagation layers consider a 
cross-graph attention matching which measures how well a node 
in one graph can be matched to one or more nodes in another graph:

\begin{equation}
  {a_{j \to i}} = \frac{{\exp \left( {{s_h}\left( {h_i^{\left( t \right)},h_j^{\left( t \right)}} \right)} \right)}}{{    \sum_{{j^{'}} \in {G_2}} {{\exp \left( {{s_h} \left( {h_i^{(t)},h_{j^{'}}^{(t)}}\right)}\right)}}}}
\end{equation}

\begin{equation}
  {\mu _{j \to i}} = {a_{j \to i}}\left( {h_i^{\left( t \right)} - h_j^{\left( t \right)}} \right)
\end{equation}

In this context, $s_h$ represents the cosine similarity, with 
$h_i^{\left( t \right)},h_j^{\left( t \right)}$ being the 
representations of node $v_i$ in graph $G_1$ and node $v_j$ in graph 
$G_2$ after the $t$-th round of propagation, respectively. 
Intuitively, the attention parameter ${a_{j \to i}}$ enables the 
instruction nodes in graph $G_1$ to pay more attention to 
semantically similar instruction nodes in graph $G_2$. Therefore, 
the attention based on the RoBERTa model's fine-tuning
can more straightforwardly focus on nodes with similar semantic.

Based on the node representations $h_i^{\left( t \right)}$ from the 
previous round of propagation, the information 
from the neighboring nodes in the current round $m_{i^{'} \to i}$ 
and the cross-graph node matching information ${u_{j \to i}}$, 
the model updates the node representations through a GRU 
(Gated Recurrent Unit):

\begin{equation}
h_{i}^{\left( {t + 1} \right)} = GRU \left( {h_{i}^{\left( t \right)},\sum\limits_{{( {i,{i^{'}}} ) \in {G_{1}}}} {{m_{{{i^{'}} \to i}}},\sum\limits_{{j \in {G_{2}}}} {{\mu _{j \to i}}} } } \right)
\end{equation}

Then, an aggregator takes the set of node representations 
$h_i^{\left( T \right)},i \in {G_1}$ as input to generate 
a graph-level representation:

\begin{equation}
{h_{{G_1}}} = MLP\left( {\sum\limits_i {\sigma ( {ML{P_{gate}}( {h_i^{\left( T \right)}} )} ) \otimes MLP( {h_i^{( T )}} )} } \right) 
\end{equation}
For each node, its representation transformed by an MLP 
is aggregated with the representations of other nodes through 
a weighted sum. This weighted sum uses weights or attention 
scores to determine the contribution of different nodes. 
Additionally, gating vectors can be used to modulate the 
importance of each node. This process helps to filter out 
irrelevant information, extract important node features, 
and better capture the structural information in the graph. 
The computation of the graph representation of ${G_2}$ is identical. 
The similarity score of the graphs can be calculated using 
standard vector space similarity, $s = {f_s}\left( {{h_{{G_1}}},{h_{{G_2}}}} \right)$.

We employ pairwise training and optimize using gradient descent 
algorithms. The pairwise loss function we used is:

\begin{equation}
{L_{pair}} = \max \left\{ {0,\gamma  - t\left( {1 - d\left( {{G_1},{G_2}} \right)} \right)} \right\}
\end{equation}

where $t=1$ if the pair is a similar function, otherwise $t=-1$. 
$\gamma  > 0$ is a margin parameter, and $d\left( {{G_1},{G_2}} \right) = {\left\| {{h_{{G_1}}} - {h_{{G_2}}}} \right\|^2}$ 
represents the Euclidean distance. Our focus is not on designing 
new networks for feature extraction but rather on how to better 
utilize suitable networks for BCSD, specifically for the binary 
graph representation we have built.



\section{IMPLEMENTATION}
\label{sec:implementation}

We leverage IDA Pro 7.7 \cite{Idapro} into disassembling binaries 
and write scripts with IDAPython \cite{idapython} that can extract 
Microcode with the information needed such as def-use 
lists and operand types. We lift the binary opting for the maturity level `MMAT\_LOCOPT' 
because the
instructions at this optimization level are relatively concise,
and the lower two maturity levels could not be successfully
exported via scripts. Additionally, we do not choose a higher
optimization level to avoid increased difficulty in node representation 
learning due to overly optimized instructions. This disassembling part is finished 
on a server running Windows 10 with Intel Xeon Silver 4210 
CPU @ 2.20GHz and 32GB RAM. We implement RoBERTa \cite{liu2019roberta} and graph matching network (GMN) \cite{li2019graph} 
using Transformers \cite{wolf2020transformers}, NetworkX \cite{hagberg2008exploring}, 
Tensorflow \cite{tensorflow2015-whitepaper} based on Python 3.7.15. We run these parts of the experiments on a 
Linux server running Ubuntu 22.04.2, with an Intel Xeon 
Platinum 8362 CPU @ 2.80GHz, 251GB RAM and one NVIDIA RTX3090 GPU.

\textbf{Hyper-parameters.} For the vocabulary generation, 
we consider the tokens which occur more than 10 times in our training dataset. 
The tokens not in our vocabulary will be seen as its operand type.
In the RoBERTa model, the node embedding dimension is 768. 
In GMN, the number of propagation layers is 10, the aggregation 
type is `sum' and the dimension of node hidden embedding is 128. 
In the structure training, the learning rate is 0.001 and the batch size is 20.

\section{EVALUATION}
\label{sec:evaluation}

\subsection{Experiment Setup}

\subsubsection{Datasets} We evaluate our model on the dataset based on that from 
\cite{marcelli2022machine} due to the disassembler version. 
The dataset consists of 24 libraries 
compiled from seven open-source projects: ClamAV, Curl, 
Nmap, OpenSSL, Unrar, Z3 and Zlib. Each library is compiled 
using two compilers (GCC, Clang) with four versions each, 
for 5 architecture combinations (x64, x86, ARM-32bit, ARM-64bit and 
MIPS-32bit) and 5 optimization levels (O0, O1, O2, O3, Os).

\subsubsection{Metrics} 

We select two commonly used metrics to measure 
ranking accuracy: MRR (Mean Reciprocal Rank) and Recall@K \cite{marcelli2022machine}. 
Recall@K calculates the proportion of truly similar items 
retrieved within the top K items. MRR measures how often a 
similar item appears at the top of a ranking list, with a 
value closer to 1 indicating better performance of the model 
in calculating similarity. MRR is computed by:

\begin{equation}
    MRR = \frac{1}{{\left\| Q \right\|}}\sum\limits_{q = 1}^Q {\frac{1}{{ran{k_q}}}} 
\end{equation}

where $Q$ represents the set of all query functions, and 
$rank_q$ represents the rank of the ground truth function for the $q$th query function.

\subsubsection{Baselines} We choose the following state-of-the-art (SOTA)
methods for comparison:

\begin{itemize}
  \item Zeek \cite{shalev2018binary}: Computes hash values for 
  instructions with data 
  dependence as feature vectors of basic blocks, 
  and then applies a two-layer fully connected neural 
  network to learn cross-architecture function similarity.
  \item SAFE \cite{massarelli2019safe}: Utilizes a seq2seq model-based NLP encoder, 
  using a self-attention sentence encoder to embed assembly code.
  \item GMN \cite{li2019graph}: Uses a bag-of-words model based on opcodes 
  as the feature of basic blocks, and then utilizing GMN 
  to extract function-level semantic. It is proven to outperform 
  concurrent work in \cite{marcelli2022machine}.
  \item jTrans \cite{wang2022jtrans}: Based on the Transformer model, 
  it extracts token embeddings from normalized Instructions 
  and adds position embeddings to better learn jump-related 
  information in instructions.
  \item Trex \cite{pei2020trex}: Besides the assembly instructions, 
  it tracks the corresponding register values during the execution 
  process, combining with transfer learning to extract the 
  semantic of function execution.
  \item VulHawk \cite{luo2023vulhawk}: In addition to using the RoBERTa model for embedding 
  node information, it also determines binary optimization levels 
  and compilers from an entropy perspective. 
\end{itemize}

\subsubsection{Evaluation setup} Our evaluation setup is presented as follows: 
Section \ref{sec:effectiveness} showcases the 
effectiveness of \textsc{StrTune} on tasks related to similar function retrieval. 
Section \ref{sec:efficiency} illustrates the computational time of each part 
of \textsc{StrTune} and efficiency of it. Section \ref{sec:ablation} delineates the 
contribution of each component of \textsc{StrTune}, 
accompanied by visual representations. 
Finally, Section \ref{sec:vulnerability} demonstrates the 
practicality of \textsc{StrTune} in real-world vulnerability search.

\subsection{Effectiveness}
\label{sec:effectiveness}

We use four different tasks to evaluate our work: 
(1) XO (Cross-Optimization): Function pairs have different 
optimizations but use the same compiler, compiler version, and 
architecture. (2) XC (Cross-Compiler): Function pairs have different 
compilers, compiler versions, and optimizations, but use the same 
architecture and bitness. (3) XA (Cross-Architecture): Function 
pairs have different architectures and bitness, but use the same 
compiler, compiler version, and optimizations. (4) XM (Mixed): 
Function pairs come from any combination of architecture, bitness, 
compiler, compiler version, and optimization.

For XO, XA and XM, we select 1,000 functions as query functions 
and randomly pick 100 and 1,000 functions that match the 
respective compilation setting variances as the function pool. 
Since jTrans is Binary Code Similarity Detection (BCSD) approaches set for x64, we 
set additional tasks of XO and XC function retrieval tasks on x64. 
For x64-XO task, we set the pool size to 100, and for x64-XC task, we 
set the pool size to 100 and 1,000.

\begin{table*}[!t]
  \renewcommand{\arraystretch}{1.2}
  \centering
  \caption{Results of our experiments on the tasks for XA, XO, and XM containing datasets of all compilation configurations 
  for poolsize=100 and poolsize=1,000 respectively. Tasks for XO, XC on the x64 datasets are denoted as x64-XO and x64-XC.
  We set poolsize=100 for x64-XO and poolsize=100/1,000 for x64-XC. The metrics are Recall@1/MRR10.}
  \label{tab01}
  \resizebox{\textwidth}{!}{
  \begin{tabular}{ccccccccccc}
    \toprule
                      & \multicolumn{5}{c}{ poolsize=100 }              & \multicolumn{4}{c}{poolsize=1,000}                 \\\cmidrule(r){2-6}    \cmidrule(r){7-10}                            
    \multirow{-2}{*}{ } & \multicolumn{1}{c}{XA} & \multicolumn{1}{c}{XO} & \multicolumn{1}{c}{XM} & \multicolumn{1}{c}{x64-XO} & \multicolumn{1}{c}{x64-XC} & \multicolumn{1}{c}{XA} & \multicolumn{1}{c}{XO} &  \multicolumn{1}{c}{XM} & \multicolumn{1}{c}{x64-XC} \\\midrule
    SAFE                &     0.098/0.219       &  0.160/0.300    &   0.110/0.225    &  0.180/0.327    &  0.124/0.253    &  0.019/0.045     &  0.030/0.074    &     0.015/0.034    &      0.013/0.040         \\\midrule
    Zeek                 &      0.257/0.416     &  0.281/0.428   &  0.212/0.346  &  0.258/0.419     &  0.193/0.338    &  0.052/0.111   &  0.087/0.164   &     0.045/0.091   &       0.045/0.090            \\\midrule
    GMN                 &     0.658/0.766     &  0.589/0.708 &   0.465/0.601 &  0.684/0.788   &  0.422/0.568     &  0.450/0.580          &  0.472/0.561   &     0.276/0.383     &     0.169/0.258           \\\midrule
    VulHawk             &   0.113/0.381      &    0.094/0.360  &   0.103/0.348  &  0.328/0.579   &   0.268/0.539  &         -           &         -       &          -          &        -              \\\midrule
    Trex                &     0.013/0.046     &  0.675/0.744 &   0.160/0.206   &  0.792/0.842   &  0.623/0.730     &  0.002/0.004          &  0.527/0.579   &     0.110/0.131     &     0.299/0.409           \\\midrule 
    jTrans             &     -           &  -              &     -           &  0.677/0.749   &  0.562/0.659    &      -              &  -            &         -            &        0.314/0.396         \\\midrule
    \textsc{StrTune}            &    \textbf{0.836/0.897}    &  \textbf{0.766/0.837}     &  \textbf{0.625/0.729}  &  \textbf{0.836/0.882}    &  \textbf{0.648/0.754}     &  \textbf{0.665/0.730}         &  \textbf{0.537/0.620}    &     \textbf{0.409/0.502}    &       \textbf{0.363/0.473}       \\
    \bottomrule
    \vspace{-1.5ex}                                                                                     
    \end{tabular}}
    \textbf{Note}: The "-" for jTrans is due 
      to the fact that the model itself only 
      analyzes x64 binaries. The "-" for VulHawk 
      is due to the 24-hour time limit, 
      within which it still cannot output all results.
\end{table*}

Table \ref{tab01} displays the results of the models tested in various tasks. 
The results indicate that \textsc{StrTune} outperforms other baselines 
in all tasks, achieving the highest Recall@1 and MRR@10. 
Specifically, in the x64-XO, x64-XC (poolsize=100), and x64-XC (poolsize=1,000) tasks, 
\textsc{StrTune} demonstrates Recall@1 improvements of 23.5\%, 15.3\%, and 15.6\% 
compared to one of the latest models, jTrans. Compared to the earlier work SAFE and Zeek, 
\textsc{StrTune} shows approximately two and three times improvements respectively in Recall@1. 
Compared to GMN, \textsc{StrTune} showcases improvements of 
47.8\%, 13.8\%, 48.2\%, and 114.8\% in the XA, XO, XM, and x64-XC (poolsize=1,000) tasks, 
suggesting that directly using CFG as a function graph representation might not be appropriate, 
and our proposed graph representation can better reflect the features of functions.

\begin{figure*}[!t]
	\centering
	\begin{subfigure}{0.27\linewidth}
    		\centering
    		\includegraphics[width=0.9\linewidth]{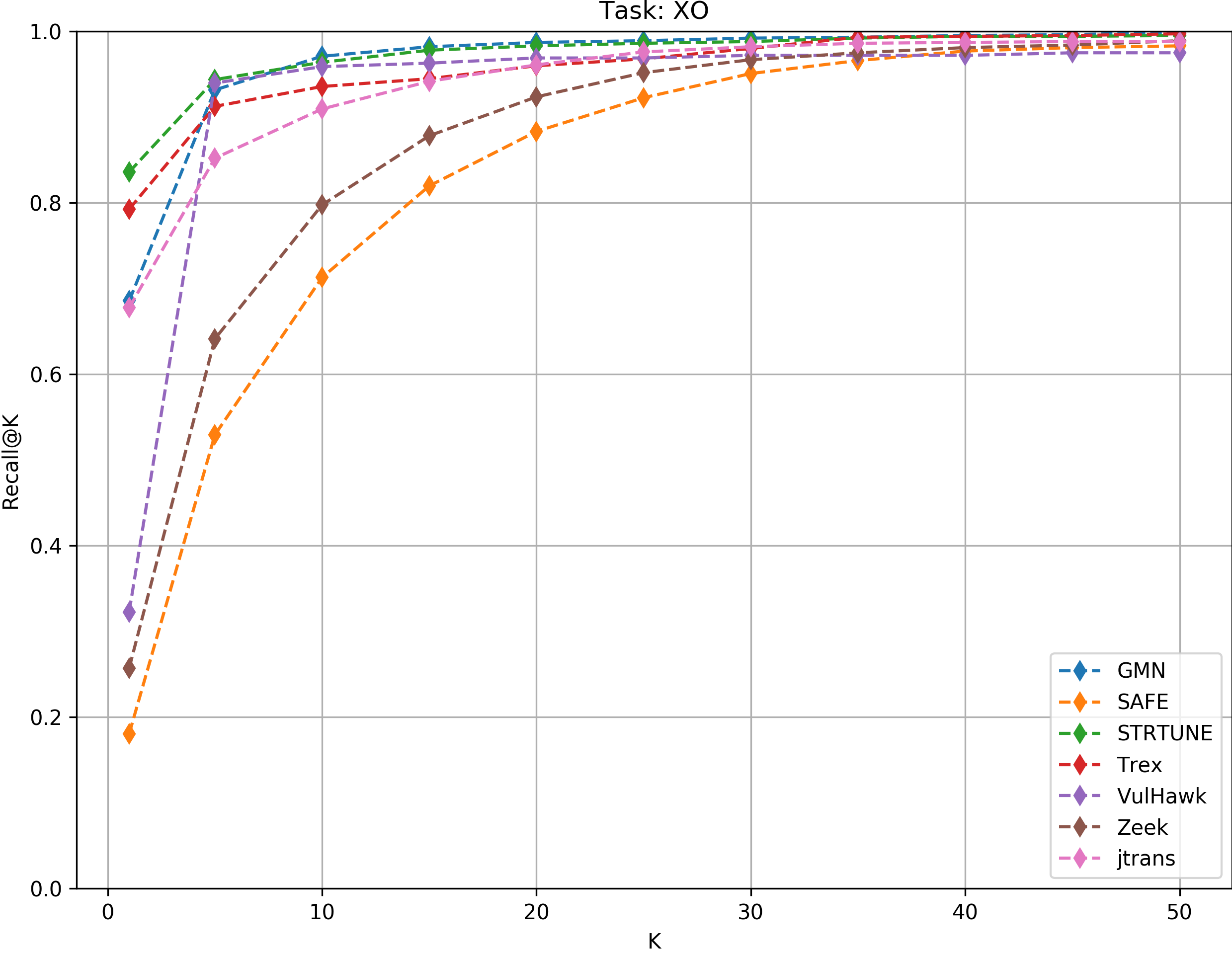}
    		\caption{x64-XO}
    		\label{4Recall}
	\end{subfigure}
		\centering
	\begin{subfigure}{0.27\linewidth}
                \centering
    		\includegraphics[width=0.9\linewidth]{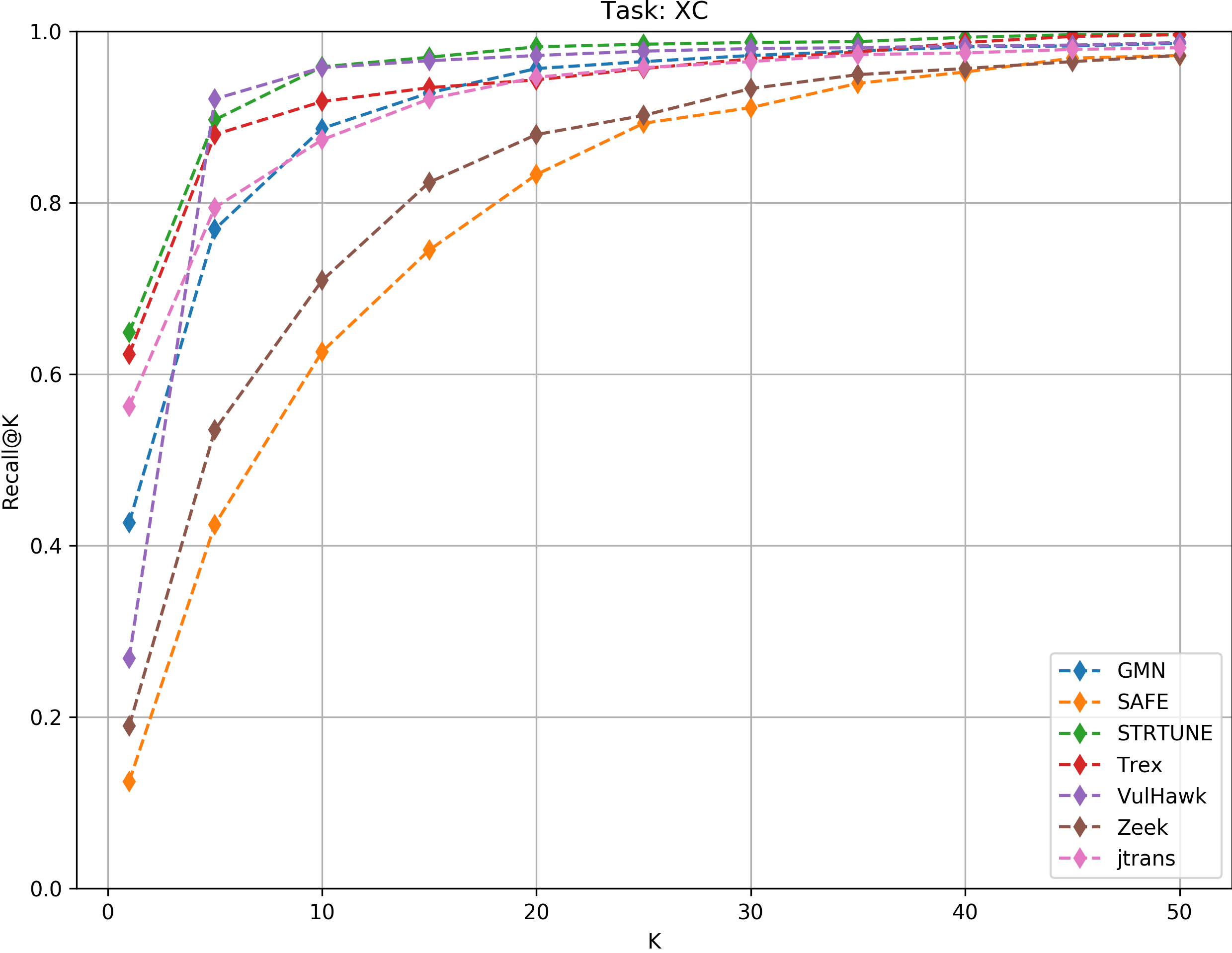}
    		\caption{x64-XC (poolsize=100)}
    		\label{5Recall}
	\end{subfigure}
        \centering
	\begin{subfigure}{0.27\linewidth}
    		\centering
    		\includegraphics[width=0.9\linewidth]{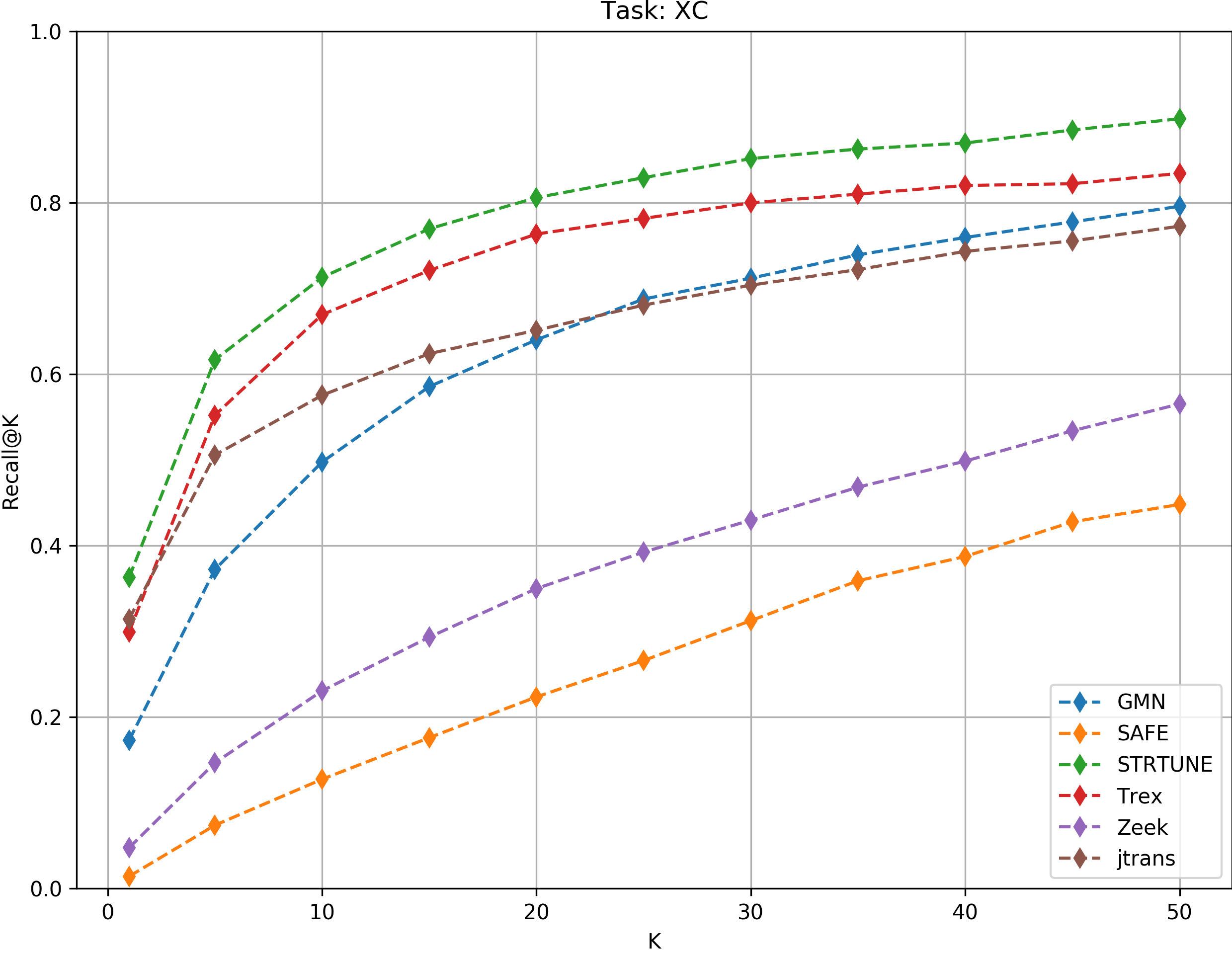}
    		\caption{x64-XC (poolsize=1,000)}
    		\label{6Recall}
	\end{subfigure}
        \centering
	\begin{subfigure}{0.27\linewidth}
    		\centering
    		\includegraphics[width=0.9\linewidth]{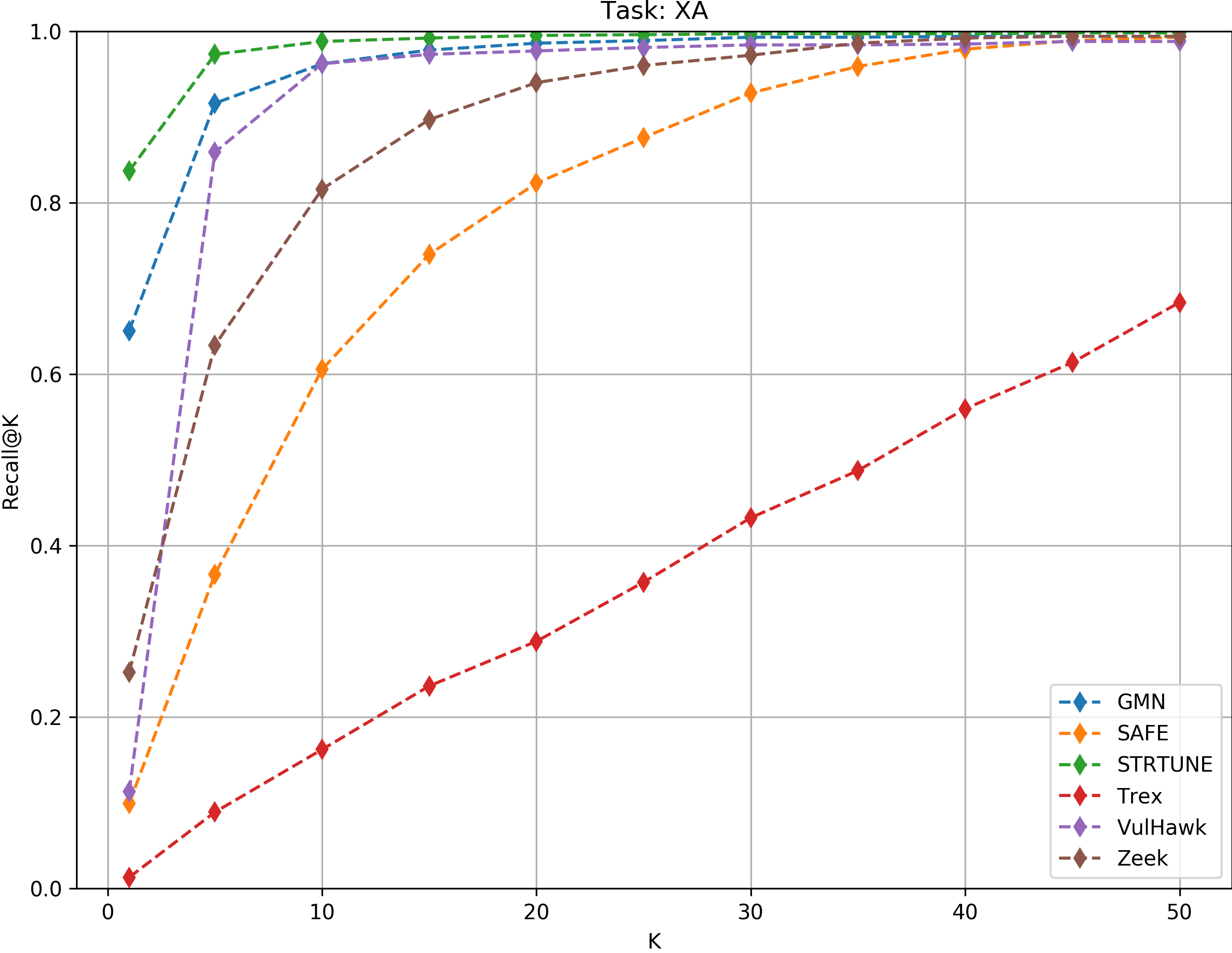}
    		\caption{XA}
    		\label{7XA}
	\end{subfigure}
        \centering
	\begin{subfigure}{0.27\linewidth}
    		\centering
    		\includegraphics[width=0.9\linewidth]{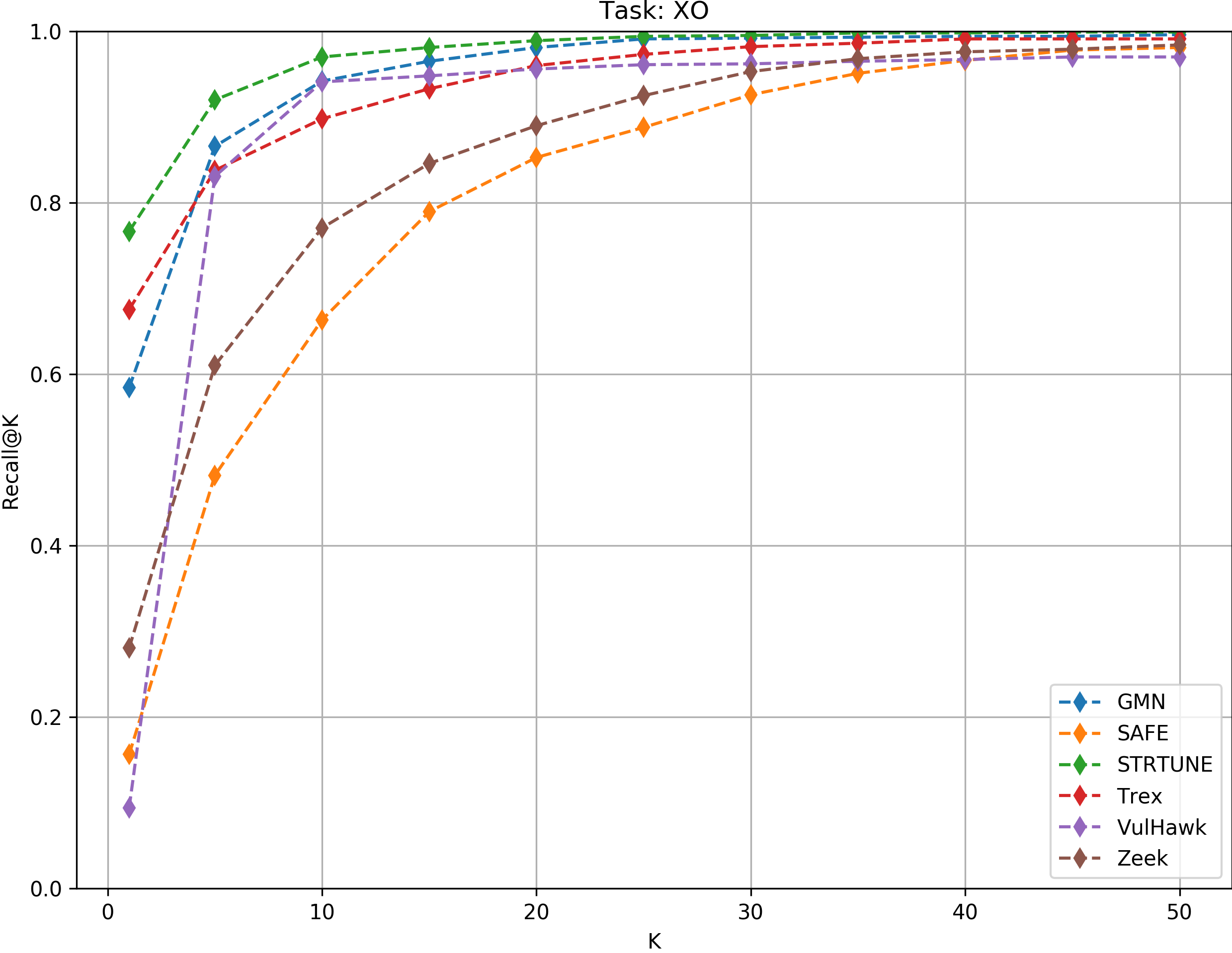}
    		\caption{XO}
    		\label{7XO}
	\end{subfigure}
        \centering
 	\begin{subfigure}{0.27\linewidth}
    		\centering
    		\includegraphics[width=0.9\linewidth]{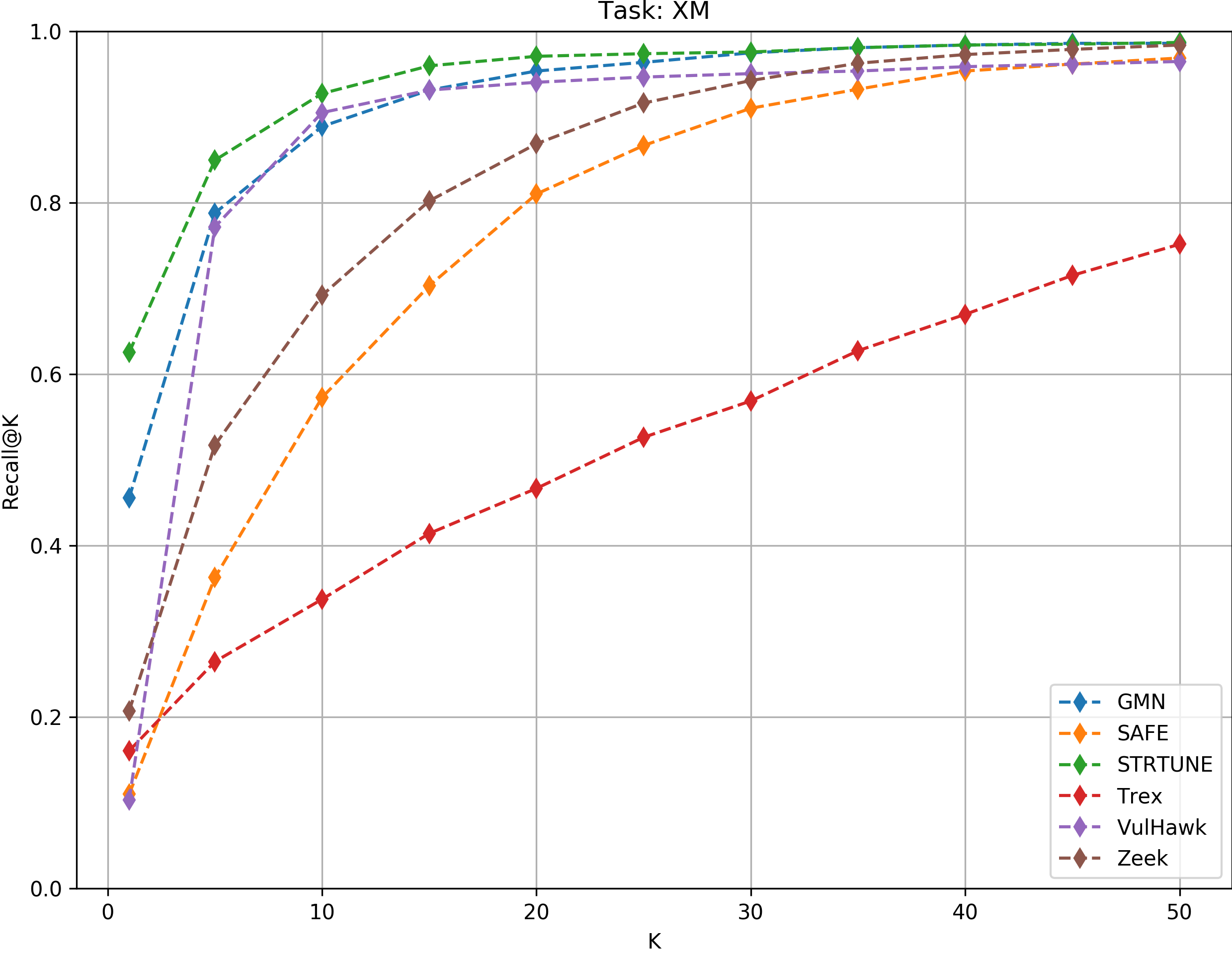}
    		\caption{XM}
    		\label{7XM}
	\end{subfigure}

   \caption{Recall on different K for tasks.}
  \label{recall}
\end{figure*}

Compared to VulHawk, 
our model shows a recall improvement of 2.54 times in Recall@1 and a 2.52 times increase in MRR@10. As shown in Figure \ref{4Recall} and Figure \ref{5Recall}, VulHawk's Recall@5 is slightly higher than \textsc{StrTune}.
However, as the value of K increases, its performance is not as good as ours. 
Its accuracy depends on a comprehensive training dataset. VulHawk mainly uses the dataset in respect of vulnerability, while the distribution of our dataset may differ from theirs. This could cause VulHawk to lack in robustness when the predicted compilation configuration is incorrect, which directly impacts the results.

Figure \ref{recall} illustrates the Recall@K curves of models concerning 
different values of K. 
\textsc{StrTune} ranks high in both Recall@1 and convergence speed. 
In the tasks of poolsize=100, \textsc{StrTune} essentially converges to nearly 1 at Recall@15, 
followed by GMN and jTrans. However, when poolsize is set to 1,000, as shown in Figure \ref{6Recall},  
jTrans surpasses GMN when K is small. This might be due to the increase of poolsize, 
GNN-based GMN, compared to NLP-based jTrans, is more prone to erroneously matching basic blocks 
that are semantically different but syntactically similar. In other words, 
basic blocks with different computational content may exhibit the same 
combinations of opcodes, leading GMN to mistakenly identify them as similar nodes.
Figure \ref{7XA}, \ref{7XO} and \ref{7XM} show the exact advantages of \textsc{StrTune} 
over baselines.

\begin{tcolorbox}[top=2.5pt, bottom=2.5pt, left=2.5pt, right=2.5pt]
\textbf{Summary:} \textsc{StrTune} ranks the first in all BCSD tasks, 
affirming its effectiveness in addressing cross-architecture, 
cross-optimization level, and cross-compiler challenges.
\end{tcolorbox}

\subsection{Runtime Efficiency}
\label{sec:efficiency}

\begin{figure}[!t]
  \centering
  \includegraphics[width=0.25\textwidth]{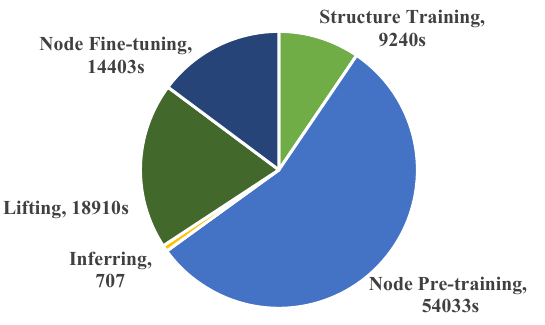}
  \caption{Efficiency of \textsc{StrTune} for execution of each part. Lifting and Inferring denote the average time taken for 100,000 pairs of functions under three rounds of execution. The time unit is seconds.}
  \label{part}
\end{figure}



\begin{figure}[!t]
  \centering
  \includegraphics[width=0.3\textwidth]{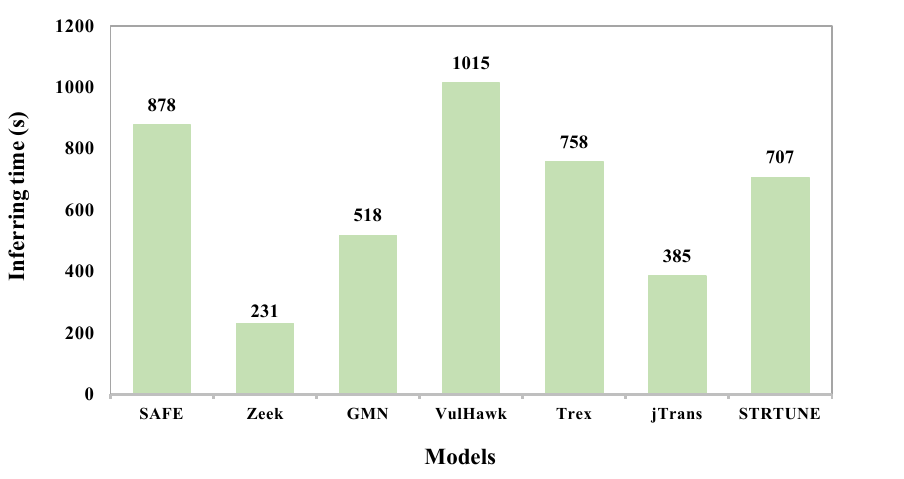}
  \caption{Average inferring time compared with baselines for 100,000 pairs of functions under three rounds of execution.}
  \label{time}
\end{figure}

We present the efficiency of \textsc{StrTune} at 
each stage in Figure \ref{part}. 
`Node Pre-training' represents the pre-training time 
of the RoBERTa model on given slices. 
`Node Fine-tuning' signifies the fine-tuning time of our RoBERTa model.
These two durations related to node representation 
are one-time efforts, hence slightly longer time can be accepted.
`Structure Training' denotes the time for 
training the model for 10 epochs. As for lifting and inferring time, 
we consider the average time tested for x64 dataset, 
totaling 100,000 pairs of functions under three rounds of execution. 
It is worth noting that due to RoBERTa 
training process where embeddings for slices
are already stored in the database, the model's 
time consumption during training and inferring stages is less than expected.

Figure \ref{time} shows average inferring time 
compared with baselines for 100,000 pairs of functions under three rounds of execution.
It can be observed 
that while Zeek consumes the least time, its 
detection accuracy is comparatively lower, which is not to our expectation. 
VulHawk consumes the most time because its 
provided interface can only compute the 
similarity between all function pairs 
given two binaries. In our one-to-many 
testing scenario, part of computation in 
VulHawk are redundant. An appropriate interface
might reduce its inferring time. \textsc{StrTune}'s 
time falls within the medium range, being 36\% 
more time-consuming than GMN. This is mainly 
attributed to the time taken for RoBERTa embedding of untreated slices. 
Due to the improved performance of \textsc{StrTune} compared to GMN, 
the additional time required is acceptable.

\begin{tcolorbox}[top=2.5pt, bottom=2.5pt, left=2.5pt, right=2.5pt]
\textbf{Summary:} \textsc{StrTune} 
maintains high precision while incurring comparable 
time overhead contrast to baselines.
\end{tcolorbox}

  \begin{table}[!t]
    \scriptsize
    \renewcommand{\arraystretch}{1}
    \centering
    \caption{Results of the ablation study on the tasks for XA, XO, XC and XM with poolsize set to 100. The metrics are Recall@1/MRR10.}
    \label{tab04}
    \resizebox{0.48\textwidth}{!}{
    \begin{tabular}{cccccc}
      \toprule
                        & XA           &      XO & XC & XM          \\\midrule                          
     w/o Data Parallel      &     0.828/0.876       &  0.757/0.813    &   0.577/0.694    &  0.616/0.706       \\\midrule
    w/o Data Dependence      &     0.829/0.881       &  0.759/0.818    &   0.576/0.689    &  0.617/0.711        \\\midrule
      w/o Jump      &     0.832/0.886       &  0.761/0.823    &   0.583/0.704    &  0.620/0.716        \\\midrule
        w/o Sequential     &     0.833/0.891       &  0.764/0.828    &   0.580/0.699    &  0.621/0.721        \\\midrule
      w/o Fine-tuning      &      0.644/0.725     &  0.619/0.698   &  0.416/0.526     &  0.485/0.593        \\\midrule
      w/o Slicing             &     0.493/0.642    &    0.473/0.610   &    0.336/0.495    &  0.371/0.527     \\\midrule
      w/o Attention Coefficient  &       0.779/0.849     &   0.728/0.806    &   0.569/0.679    &   0.621/0.725                 \\\midrule
      \textsc{StrTune}            &    \textbf{0.836/0.897}    &  \textbf{0.766/0.837}        &  \textbf{0.585/0.709}    &  \textbf{0.625/0.726}        \\
      \bottomrule
      \end{tabular}}
  \end{table}

\begin{figure}[!t]
	\centering
	\begin{subfigure}{0.4\linewidth}
    		\centering
    		\includegraphics[width=0.9\linewidth]{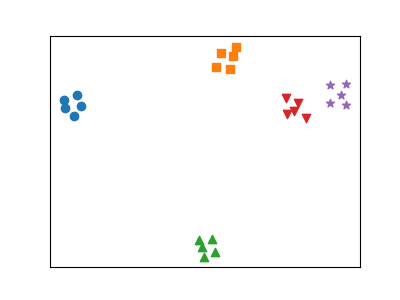}
    		\caption{Code slices embedded by the fine-tuned RoBERTa of \textsc{StrTune}.}
    		\label{umap_StrTune}
	\end{subfigure}
        \centering
	\begin{subfigure}{0.4\linewidth}
    		\centering
    		\includegraphics[width=0.9\linewidth]{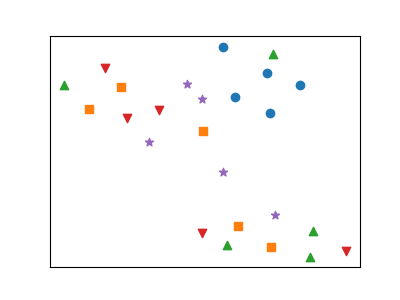}
    		\caption{Code slices embedded by the `roberta-base' model used for NLP embedding.}
    		\label{umap_roberta}
	\end{subfigure}

     \caption{Visualization of clustering of code slices with different semantic. 
    We select five syntactically different slices from different compilation configurations for five diverse computational contents 
    and use Uniform Manifold Approximation and Projection (UMAP) for dimensionality reduction.}
    \label{umap}
\end{figure}

\subsection{Ablation Study}
\label{sec:ablation}

We analyze the contributions of each part of our model through ablation analysis. 
We mainly test the models on the XO/XA/XC/XM tasks. We select the 
following input representation and embedding models as comparison:

\begin{itemize}
  \item w/o Data Parallel/Data Dependence/Jump/Sequential: We individually removed each flow type to observe its contribution to the overall performance, and the initial features of the removed flow type were set to 1. Nodes remain as code slices 
  in the graph, and the node embedding method remains unchanged.
  \item w/o Fine-tuning: We change the node embedding method to an unfine-tuned RoBERTa model.
  \item w/o Slicing: The input graph is obtained directly from Microcode and nodes 
  are basic blocks. The method of node embedding remains the same, while edges retain 
  only two types: control flow and data dependence. We set this input to verify the effectiveness of 
  our proposed graph presentation.
  \item w/o Attention Coefficient: We transform the structural learning GMN into a regular Graph Neural Network (GNN) model.
\end{itemize}

As shown in Table \ref{tab04}, each component of \textsc{StrTune} 
contributes to the improvement of BCSD accuracy. 
The results of w/o Slicing are the poorest, indicating 
that regarding basic blocks as graph nodes are 
inadequate. In contrast, our proposed graph representation 
enhances Recall@1 by 69\%, validating the effectiveness 
of node slicing based on data dependence. Model w/o Fine-tuning has a reduction 
of around 0.2 in Recall@1 when compared. 
This demonstrates that pre-training RoBERTa alone cannot 
effectively learn slice representations with 
similar semantic but different syntax, highlighting 
the importance of fine-tuning step. With regard to structure 
learning, employing GMN results in a 7.3\% improvement 
compared to GNN, indicating that attention coefficients 
are beneficial for graph representation learning. 
In fact, \textsc{StrTune} focuses on the cross-graph node matching, 
which can be amplified by the attention coefficients through 
several rounds of node propagations, promoting efficiency in BCSD. With respect to flow type, every type of flow contributes to our overall performance, with data parallel contributing the most to the effectiveness, followed by data dependence. Data parallel and data dependence enrich the data-level relationships between slices, thereby enhancing the robustness of \textsc{StrTune}.

We can show the effectiveness of the fine-tuned RoBERTa model 
on representing code slices through visualization. We utilized Uniform 
Manifold Approximation and Projection (UMAP) to perform non-linear 
dimensionality reduction on code slice embeddings. 
Figure \ref{umap} presents clustering of several code slices. 
Points with the same color represent different slices with the same semantic
(from the same source code, for example, $while(o \to nb\_stream\_maps > v7)$ shown before).
It is evident that our fine-tuned RoBERTa model can generate embeddings that 
bring semantically similar slices closer together. 
In contrast, the SOTA NLP model `roberta-base' when used to embed code slices, 
cannot cluster them very effectively.
This improvement provides a strong foundation for structure-level learning for functions.

\begin{figure*}[!t]
	\centering
	\begin{subfigure}{0.39\linewidth}
    		\centering
    		\includegraphics[width=0.9\linewidth]{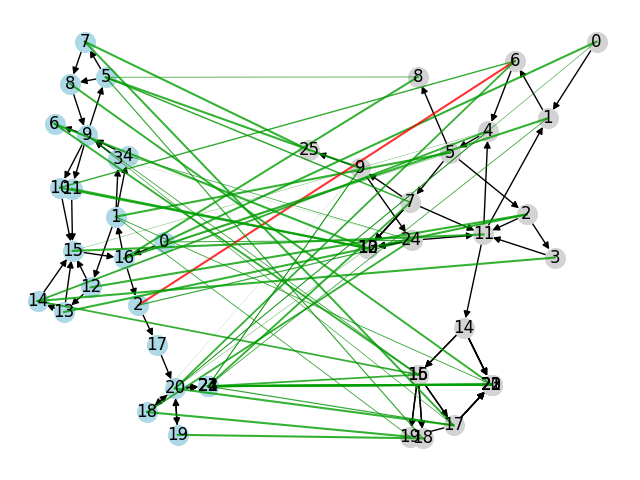}
    		\caption{Attention after 1 round of propagation.}
    		\label{attention 1}
	\end{subfigure}
        \centering
	\begin{subfigure}{0.39\linewidth}
    		\centering
    		\includegraphics[width=0.9\linewidth]{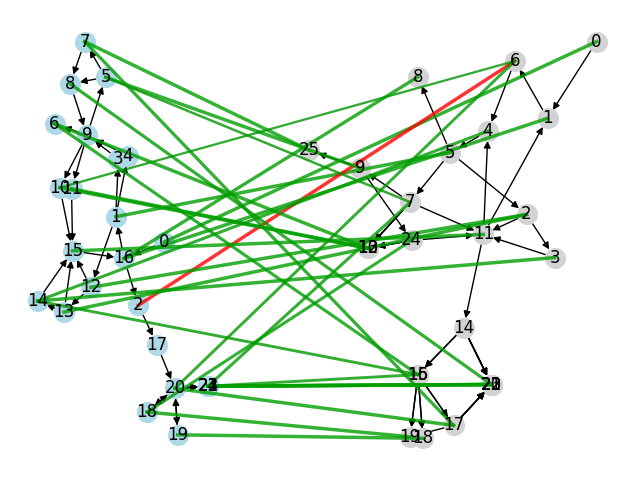}
    		\caption{Attention after 10 rounds of propagation.}
    		\label{attention 2}
	\end{subfigure}

  \caption{Visualization of attention coefficients 
  between a pair of similar functions during inference. 
  The intensity and thickness of the line correspond 
  to the attention coefficient, with darker and thicker lines 
  indicating a higher degree of similarity between the connected slices.}
  \label{attention}
\end{figure*}

We also visualize the attention coefficients
between a pair of real-world function graphs under different rounds of propagation layers
in GMN in Figure \ref{attention}. To observe the changes in attention weights more clearly, we ignore the distinction between different flow types and primarily focus on how the attention mechanism can concentrate on similar slices between the constructed graphs. Specifically, Slice 2 in blue and Slice 6 in gray both refer to the source code ${\rm{void* dst = (uint8\_t*)o + po}} \to {\rm{u}}{\rm{.off}}$. The network focuses on the edges connecting them and the similarity between the corresponding slices. Specifically, Slice 2 in blue is connected to its preceding node Slice 16, along with the jump flow between them, and Slice 6 in gray is connected to its preceding node Slice 1, and the corresponding jump flow. Since Slice 16 and Slice 1 are semantically equivalent, in the following rounds, the network will assign higher attention coefficients to Slice 2 and Slice 6, as shown in Figure \ref{attention 2}. In this case, when the similarity score is calculated at the end, this pair of functions will receive a higher similarity score.

\begin{tcolorbox}[top=2.5pt, bottom=2.5pt, left=2.5pt, right=2.5pt]
\textbf{Summary:} Each component 
of \textsc{StrTune} contributes positively to the overall performance 
and slicing based on data dependence contributes the most to the precision.
\end{tcolorbox}

\subsection{Real-world Vulnerability Search}
\label{sec:vulnerability}

\begin{table*}[!t]
  \scriptsize
  \renewcommand{\arraystretch}{1}
  \centering
  \caption{Results of real-world vulnerability searching experiments in TP-Link. The four numbers mean the ranks 
  of ground truth functions when the query function is compiled for the architectures of 
  x86, x64, ARM-32bit and MIPS-32bit respectively.}
  \label{tab05}
  \resizebox{\textwidth}{!}{
    \begin{tabular}{cccccccc}
      \toprule
                     &    CVE-2016-2182       & CVE-2016-0797  &    CVE-2016-0797&CVE-2016-2105&CVE-2019-1563&CVE-2016-0798&CVE-2016-2176           \\                          
      \multirow{-2}{*}{ CVE   } &(BN\_bn2dec) &(BN\_dec2bn) & (BN\_hex2bn) & (EVP\_EncodeUpdate)& (PKCS7\_dataDecode)&(SRP\_VBASE\_get\_by\_user)&(X509\_NAME\_oneline)\\\midrule
      SAFE                &     118;92;62;57     &  68;77;79;44   &   88;172;63;70 &61;333;29;93&243;52;153;58&13;27;152;5&63;16;18;29  \\\midrule
      Zeek                 &     86;71;51;102    &  5;1;5;51 &  434;2;319;281&10;4;23;\textbf{1}&\textbf{3;1;}1;3&11;94;14;19&3;1;1;4\\\midrule
      GMN                 &    9;35;104;39   & 1;1;1;2 &  1;1;1;1&79;6;2;\textbf{1}&17;92;36;1&1;1;2;6&29;24;16;61 \\\midrule
      \textsc{StrTune}            &   \textbf{1;1;4;1}   &  \textbf{1;1;1;1}    &  \textbf{1;1;1;1}  &\textbf{1;1;1;}4&9;3;\textbf{1;1}&\textbf{1;1;1;1}& \textbf{1;1;1;1} \\
      \bottomrule                                                                                      
    \end{tabular}}
\end{table*}

\begin{table}[!t]
  \scriptsize
  \renewcommand{\arraystretch}{0.8}
  \centering
  \caption{Results of real-world vulnerability searching experiments in NetGear. The four numbers mean the ranks 
  of ground truth functions when the query function is compiled for the architectures of 
  x86, x64, ARM-32bit and MIPS-32bit respectively.}
  \label{tab06}
  \resizebox{0.48\textwidth}{!}{
    \begin{tabular}{cccc}
      \toprule
                     &    CVE-2016-2182       & CVE-2016-6303   &    CVE-2019-1563            \\                          
      \multirow{-2}{*}{ CVE   } &(BN\_bn2dec) &(MDC2\_Update) & (PKCS7\_dataDecode) \\\midrule
      SAFE                &     42;42;128;17      &  183;110;139;384   &   99;3;28;26    \\\midrule
      Zeek                 &     35;59;35;22    &  252;62;5;345  &  3;7;15;10 \\\midrule
      GMN                 &    16;17;10;53    & 2;3;1;7 &   108;1;\textbf{1;}2  \\\midrule
      VulHawk             &     1;11;5;3     &  1;2;1;\textbf{1}  &  1;1;\textbf{1;}123 \\\midrule
      \textsc{StrTune}            &   \textbf{1;1;1;1}    &  \textbf{1;1;1;}2     &  \textbf{1;1;}6;\textbf{1}    \\
      \bottomrule
  \end{tabular}}
\end{table}

In order to test the \textsc{StrTune}'s generalizability in real-world scenarios, 
we conduct a real-world vulnerability search on \textsc{StrTune} and some 
baselines. We select six CVEs (Common Vulnerabilities and Exposures) 
from OpenSSL 1.0.2d, which include eight functions. The 
functions are compiled for four architectures 
and serving as query functions. We choose NetGear R7000 (ARM-32) 
from NetGear and TP-Link Deco M4 (MIPS-32) from TP-Link as the function repositories. 
Three query functions are present in the NetGear R7000, and seven query 
functions in the TP-Link Deco M4. For each firmware, we evaluate 
the ranking of the corresponding functions in four architectures.

Table \ref{tab05} and Table \ref{tab06} respectively show the search 
results of the CVE functions in the two firmwares. Each score represents 
the ranking of the corresponding ground truth function in the function repository, 
compiled for x86, x64, ARM-32bit, 
and MIPS-32bit.

SAFE and Zeek exhibit relatively low accuracy 
in both firmware detection tasks, even ranking 
the ground truth function beyond 100 for more than half CVE instances. 
In contrast, 
GMN achieves higher accuracy, almost 
performing as well as \textsc{StrTune} in searching for 
CVE-2016-0797 and CVE-2016-0797, but performing 
poorly for other CVEs. \textsc{StrTune} enables 
highly accurate matching, achieving a top 
ranking for five CVEs across four architectures, 
with the lowest search result ranking being 9.
\textsc{StrTune} surpasses the searching accuracy over other baselines, 
which illustrates the high robustness of slice-based representation. 
We interpret that \textsc{StrTune} can
effectively improve vulnerability search tasks in real-world scenarios, 
significantly reducing the cost of manual function comparison for users.

We analyze false positives in the aforementioned experiments, which 
means that functions compiled from 
different source code are ranked at 1 during function searching. 
For instance, in the case of CVE-2016-2182 compiled for 
ARM-32bit and searched in TP-Link Deco M4, \textit{OBJ\_dup} ranks at the top. 
We check the node matching pattern 
of two functions during the inferring phase. Due to our retention of all call function names, 
the code slices involving calls to \textit{OPENSSL\_malloc} and 
\textit{OPENSSL\_free}, along with their associated flow types, are nearly identical. 
This prompts \textsc{StrTune} to focus on these specific local details. 
GMN tends to assign high attention coefficients to these two functions, resulting in a higher similarity score. 
In contrast, the ground truth function ranks at 4 due to the failure to match slices with similar computational contents, 
thus leading to a false negative.

Since such cases account for a significant portion of false positives and false negatives, 
We can incorporate dissimilar slices from false positives into 
the negative samples, with similar slice pairs from false negatives 
adding as positive samples for fine-tuning and then re-fine-tune RoBERTa model. The GMN could assign 
lower attention to those dissimilar slices, thus giving a low similarity score to these functions.

In another case, we identify \textit{EVP\_DigestSignInit} as a false positive for \textit{EVP\_DigestVerifyInit}. 
These two functions are used for digital signatures and are almost identical in terms of their functionality, 
due to which they are considered similar functions by \textsc{StrTune}.

\begin{tcolorbox}[top=2.5pt, bottom=2.5pt, left=2.5pt, right=2.5pt]
\textbf{Summary:} \textsc{StrTune} demonstrates high performance in 
distinguishing vulnerable functions compared to SOTA methods in real-world scenarios.
\end{tcolorbox}

\section{DISCUSSION}
\label{sec:discussion}

This section discusses the limitations of our work 
and future work.

\textit{Choice of IR.} \textsc{StrTune} captures the semantic of functions based on Microcode. 
During the acquisition phase of Microcode, 
we utilize IDA Pro for decompilation, 
and the results of our work partly depend on the accuracy of this process. 
Additionally, due to version limitations of IDA Pro, 
\textsc{StrTune} is unable to obtain Microcode of binaries complied from MIPS64, 
leading to certain analytical limitations. 
Ghidra could also be used as a tool for obtaining IR for analysis.

\textit{Model Efficiency and Enhancements.} \textsc{StrTune} 
is based on the pre-training and fine-tuning of models.
How to effectively use NLP models for encoding or adopting more 
efficient training methods is one of our future directions. Also, 
in the structure training phase, it is also possible to represent 
the relationship between nodes using the attention mechanism 
of Query, Key, Value.
Although this approach essentially involves 
increasing a certain amount of trainable parameters, 
we still consider it as one of our future work.

\textit{Hierarchical Fusion of Learning Representations.} 
Since our work trains nodes and structures separately, 
we could also use a Hierarchical Attention Network for 
learning. It mainly includes two levels of attention mechanisms, 
which we can apply to instruction-level and structure-level attention, 
respectively, to enhance the model's understanding and 
representation capabilities of the entire function composition.

\textit{Problems of same code behavior.} 
In the real-world vulnerability search, 
we discovered that functions with different names 
but the same code logic introduce some noise into the analysis. 
How to effectively remove this part of the noise is also a potential future work.

\section{RELATED WORK}
\label{sec:related work}

The related work in the domain of Binary Code Similarity Detection (BCSD) is various. 
In Asm2Vec \cite{ding2019asm2vec}, the authors base their approach on the 
word2vec model, splitting assembly instructions into 
character-level units and training character-level embeddings 
using unsupervised methods. Like the word2vec-based model, 
SAFE \cite{massarelli2019safe} employs a seq2seq model-based NLP encoder. 
Ahn et al. \cite{ahn2022practical} construct a pre-trained BERT model based 
on MLM and NSP tasks, along with fine-tuning using additional 
functions. In jTrans \cite{wang2022jtrans}, the authors extract token embeddings from the normalized 
instruction set and add position embeddings, considering 
the positional relationships of source and target tokens for 
jump instructions. Trex \cite{pei2020trex} focuses 
on the concept of Micro-trace and utilizes transfer learning 
to learn the semantic of binary functions. In recent research \cite{xu2023improving}, 
Xu et al. point out that compilation introduces instruction 
distribution bias, proposing classification importance and semantic 
importance for instructions to improve the accuracy of jTrans and Trex.

Feng et al. \cite{feng2016scalable} introduce ACFG, where attributes mainly 
fall into statistical and structural categories. Gemini \cite{xu2017neural} employs a variant 
of the Structure2vec algorithm for each node in ACFG and uses 
Siamese networks for supervised learning. Zeek \cite{shalev2018binary}, based on 
data dependence, computes hash values for sets 
of instructions with data dependence, serving as basic block 
feature vectors. Li et al. \cite{li2019graph} propose GMN, 
which uses attention mechanisms to reduce the 
impact of dissimilar nodes on the results. Yu et al. \cite{yu2020order} 
combine the BERT algorithm to extract features for each 
basic block and proposed a framework with three components. In Sem2vec \cite{wang2023sem2vec}, the authors use 
symbolic constraints as node information and aggregate 
them using RoBERTa models and some structural 
network techniques. VulHawk \cite{luo2023vulhawk}, on the other hand, determines the optimization level and compiler of binaries from an entropy perspective and proposes a binary search framework that combines 
coarse-grained search with fine-grained filtering. Recently, CLAP \cite{wang2024clap} leverages contrastive language-assembly pre-training to improve the transferability of binary code representation learning by aligning assembly code with natural language explanations. BinCola \cite{jiang2024bincola} also leverages diversity-sensitive contrastive learning to improve performance across varying compilations. CEBin \cite{wang2024cebin} combines embedding-based and comparison-based approaches to enhance accuracy, particularly for large-scale vulnerability detection. He et al. \cite{he2024code} presents a semantics-oriented graph representation to better capture binary semantic.

\section{CONCLUSION}
\label{sec:conclusion}

In this paper, we propose \textsc{StrTune}, 
a novel approach that fine-tunes the representation of code slices segmented 
based on data dependence. \textsc{StrTune} emphasizes non-interference 
in computation by introducing flow types to depict relationships 
among slices, presenting a novel graph representation for binary functions. 
We employ a Siamese 
network to fine-tune pairwise slices with 
the same computational contents. We use 
GMN for function-level features. The attention 
coefficient focuses specifically on similar nodes across 
graphs. Our experimental results demonstrate that the performance 
of \textsc{StrTune} surpasses state-of-the-art models for BCSD, proving its 
effectiveness and practicality.

\ifCLASSOPTIONcompsoc
  \section*{Acknowledgments}
\else
  \section*{Acknowledgment}
\fi

The authors would like to thank the editor and anonymous reviewers for their valuable feedback to improve the manuscript. This work is partially supported by the National Key Research and Development Program of China (No. 2021YFB3101402),
Shanghai Pujiang Program (No. 22PJ1405700), 
and Shanghai Committee of Science and Technology, China (No.23511101000).

\ifCLASSOPTIONcaptionsoff
  \newpage
\fi



\bibliographystyle{ieeetr}
\bibliography{ref}
\end{document}